\documentclass[letter,12pt]{article}
\pdfoutput=1 % if your are submitting a pdflatex (i.e. if you have
\usepackage{jheppub} % for details on the use of the package, please
\usepackage[utf8]{inputenc}

\usepackage[T1,T2A]{fontenc} % if needed

\usepackage{amsmath}

\usepackage{ upgreek }

\def\bea{\begin{eqnarray}}
\def\eea{\end{eqnarray}}

\newcommand{\Z}{{\mathbb Z}}
\newcommand{\R}{{\mathbb R}}
\newcommand{\Sp}{{\mathbb S}}
\newcommand{\e}{\epsilon}

\title{Krylov complexity in quantum field theory, and beyond }

\author[a]{ Alexander Avdoshkin,} 
\author[b]{ Anatoly Dymarsky,} 
\author[c]{Michael Smolkin}
\affiliation[a]{Department of Physics, MIT, Cambridge, MA 02139, USA\\}
\affiliation[b]{Department of Physics and Astronomy, \\ University of Kentucky,\\
506 Library Drive, \\Lexington, KY, USA 40506\\}
\affiliation[c]{The Racah Institute of Physics, The Hebrew University of Jerusalem, \\ Jerusalem 91904, Israel}
\abstract{
We study Krylov complexity  in various models of quantum field theory: free massive bosons and fermions on flat space and on spheres, holographic models, and lattice models with the UV-cutoff. In certain cases we find asymptotic behavior of Lanczos coefficients, which goes beyond previously observed universality. We confirm that in all cases the exponential growth of Krylov complexity  satisfies  the conjectural inequality, which generalizes the Maldacena-Shenker-Stanford bound on chaos.  We discuss temperature dependence of Lanczos coefficients and note that the relation between the growth of Lanczos coefficients  and chaos may only hold for the sufficiently  late, truly asymptotic regime  governed by the physics at the UV cutoff. Contrary to previous suggestions, we show scenarios when Krylov complexity in quantum field theory  behaves qualitatively differently from the holographic complexity.
}

\notoc

\begin{document} 
\maketitle
\flushbottom

\section{Introduction}
\label{sec:intro}
Dynamics in Krylov space, and the associated K(rylov) complexity, have recently emerged as a new probe of the chaotic dynamics of quantum systems. Starting from an autocorrelation function $C(t)$ of a sufficiently simple,  e.g.~local operator $A$, via recursion method one can define Lanczos coefficients $b_n$, which control and characterize the operator growth in the Krylov subspace. The original work \cite{Parker_2019} proposed 
the universal operator growth hypothesis, which connects the asymptotic behavior of $b_n$ with the type of dynamics exhibited by the underline system. Namely, for a generic physical systems without apparent or hidden symmetries $b_n$ will exhibit maximal possible growth consistent with locality,  $b_n \propto n$ for spatially extended systems in $D>1$. This hypothesis is essentially the quantum version of an earlier observation, that relates the high-frequency tail of the power spectrum $f^2(\omega)$ -- the Fourier of $C(t)$  -- with presence or lack of integrability in classical spin models \cite{elsayed2014signatures}. Similarly, the hypothesis can be understood as a statement that in typical non-integrable lattice systems the norm of nested commutators $[H,[H,\dots A]]$, or equivalently the norm of $A$ subject to Euclidean time evolution, grows with the maximal speed allowed by universal geometric constraints \cite{Avdoshkin_2020}.  There is non-trivial evidence supporting the connection between the behavior of $b_n$ and integrability/chaos, yet it does not seem to be universal. In particular a possible stronger  formulation, relating the linear growth of $b_n$ specifically to  chaotic behavior of the underlying systems is apparently wrong \cite{Dymarsky:2021bjq,Bhattacharjee_2022}. We observe that for continuous systems and local  operator $A$, Lanczos coefficients always exhibits linear growth. The situation is changed if a UV-cutoff is introduced, in which case we propose the asymptotic behavior of $b_n$ would probe  integrability or lack thereof in the lattice model underlying the UV regime.

Besides  being a probe of chaos, the Krylov complexity  has a very intriguing and non-trivial connection with the exponent controlling the growth of the Out of Time Ordered Correlator (OTOC). Namely, the original work \cite{Parker_2019} conjectured (and proved for the case of infinite temperature) that the exponent of Krylov complexity bounds the exponent of OTOC, $\lambda_{\rm OTOC}\leq \lambda_{\rm K}$. We further propose this inequality being a part of a stronger relation, which generalizes the Maldacena-Shenker-Stanford bound on chaos \cite{MSS},
\bea
\label{MSS}
\lambda_{\rm OTOC}\leq \lambda_{\rm K}\leq {2\pi \over \beta}.
\eea
In the most cases considered so far first inequality in \eqref{MSS} is non-trivial, while the second one trivially becomes an equality \cite{Avdoshkin_2020}. A proof of \eqref{MSS} in such a scenario was recently given in \cite{Gu_2022}. Below we give  examples of free massive bosons and fermions, when the first inequality is trivial, while the second one becomes non-trivial. We numerically show that in this case $ \lambda_{\rm K}\leq {2\pi \over \beta}$ is satisfied, thus providing further evidence for \eqref{MSS}. 

The third virtue of Krylov complexity is its potential connection to other measures of complexity in quantum systems, in particular, holographic complexity \cite{Susskind:2014rva,Brown:2015bva,Brown:2015lvg,Ben-Ami:2016qex,Chapman:2016hwi,Carmi:2017jqz,Belin:2021bga}. Earlier studies of K complexity, specifically, in the SYK model \cite{Jian_2021}, suggested its qualitative behavior matches that one of holographic complexity. In particular, complexity grows linearly in earlier times and then saturates at the exponential values \cite{Susskind:2014rva,Carmi:2017jqz}. We analyze K-complexity for fields compactified on spheres and show its behavior is different from the one described above. Hence, we concluded that K complexity, despite its name, in field theories can be qualitatively very different from its computational or holographic ``cousins.''\footnote{Krylov complexity draws its name from the fact that its definition matches certain axioms of complexity \cite{Parker_2019}, but there was no strong reason to expect it would match computational or holographic complexities, which characterize the whole quantum state of the system, not merely an operator growth in Krylov space. An attempt to reconcile these discrepancies is recently taken in \cite{https://doi.org/10.48550/arxiv.2202.06957}.}

The connection between K complexity and open questions of quantum dynamics make it a popular subject of study \cite{dymarsky2020quantum,PhysRevLett.124.206803,Yates2020,Yates:2021lrt,Yates:2021asz,Dymarsky:2021bjq,Noh2021,Trigueros:2021rwj,https://doi.org/10.48550/arxiv.2207.13603,Fan_2022,Kar_2022,Barbon:2019wsy,Rabinovici_2021,Rabinovici_2022,Rabinovici2022,https://doi.org/10.48550/arxiv.2109.03824,Caputa_2021,https://doi.org/10.48550/arxiv.2205.05688,PhysRevE.106.014152,Bhattacharjee_2022,https://doi.org/10.48550/arxiv.2203.14330,https://doi.org/10.48550/arxiv.2204.02250,https://doi.org/10.48550/arxiv.2205.12815,https://doi.org/10.48550/arxiv.2207.05347,H_rnedal_2022,https://doi.org/10.48550/arxiv.2208.13362,https://doi.org/10.48550/arxiv.2208.10520,https://doi.org/10.48550/arxiv.2208.08452,https://doi.org/10.48550/arxiv.2210.06815,Bhattacharjee:2022lzy,He:2022ryk,Alishahiha:2022anw,Bhattacharjee:2022qjw,Bhattacharjee:2022ave,Camargo:2022rnt,Khetrapal:2022dzy}, though with a few exceptions most of the literature is focusing on discrete models. 
In this paper we continue the study of K complexity in quantum field theory, initiated in \cite{Dymarsky:2021bjq}. There we considered only conformal field theories in flat space.  Now we consider models with mass, compact spatial support and/or UV-cutoff and find rich behavior, supporting main conclusions outlined above. Interestingly, each deformation we considered -- mass, compact space or UV-cutoff have their own imprint on Lanczos coefficients $b_n$, though universality of these imprints  is unclear.  Our paper sheds light on temperature dependence  of $b_n$  as well as the role of locality. It frames  QFT as an IR limit of some discrete system, connection the behavior of $b_n$ between different limiting cases.

 This paper is organized as follows. In section \ref{sec:prel} we give preliminaries of Lanczos coefficients and Krylov complexity. We proceed in section \ref{sec:freemassive} to consider free massive scalar and fermions to observe the effect of mass, which cases the exponent of Krylov complexity to reduce. In section \ref{sec:onS} we compactify  CFTs on a sphere to notice this results in a capped Krylov complexity, at least in certain scenarios. Section \ref{sec:TandUV} continues with the role of temperature and UV cutoff. We conclude with a discussion in section \ref{sec:discussion}.

\section{Preliminaries: dynamics in Krylov space and chaos}
\label{sec:prel}
%Cao_2021
All three formulations of the ``signature of chaos'' mentioned in the introduction: through the high-frequency behavior of the power spectrum \cite{elsayed2014signatures}
\bea
f^2(\omega)={1\over 2\pi}\int dt e^{i \omega t} C(t),
\eea
through the growth of $|A(it)|$ \cite{Avdoshkin_2020}, and through the asymptotic behavior of $b_n$ \cite{Parker_2019}, are essentially mathematically equivalent, but there are some important  subtleties. The exponential asymptotic 
\bea
f^2(\omega) \sim e^{-\omega/\omega_0}  \label{fasym}
\eea
is equivalent to a pole of $C(t)$ at $t=i/\omega_0$, which is the same as the divergence of the Frobenius norm of $e^{-\beta H/4}A(t)e^{-\beta H/4}$ at $t=i/(2\omega_0)$, 
%\footnote{Divergence of the norm of $A(t)$ at some $t=i\tau^*$ implies divergence of the norm of $e^{-\beta H/4}A(t)e^{-\beta H/4}$ at $|-it|\leq \tau^*$, see \cite{Avdoshkin_2020}.???}
as follows from the definition
\bea
\label{C}
C(t) \propto {\rm Tr}(e^{-\beta H/2}A(t) e^{-\beta H/2} A(0)).
\eea
This is equivalent to asymptotic linear growth
\bea
b_n \sim {\pi \over 2\omega_0} n \label{basym}
\eea
{\it assuming}  asymptotically $b_n$ is a smooth function of $n$. Thus \eqref{basym} implies \eqref{fasym}, but the other way around may not be true provided $b_n$ exhibits some complicated behavior, for example $b_n$ split into two branches for even and odd $n$. 
The reference \cite{Avdoshkin_2020} provided a mathematical example  of this kind,\footnote{An example of a physical system exhibiting different power law behavior of even and odd branches of $b_n$ is possibly given by the large-$q$ SYK model \cite{Bhattacharjee:2022ave}.}  when $b_n$ asymptotically split into two branches,
\bea
b_n^2 \sim \left\{\begin{array}{c} 
n^2,\quad n\,\,\,\, {\rm is\, even},\\
n,\qquad n\, \, {\rm is\, odd},
\end{array} \right.
\eea
while corresponding power spectrum exhibits asymptotic behavior $f^2\propto \omega^{-\omega}$, which is an analog of  \eqref{fasym} for 1D systems.\footnote{
For 1D systems two-point function $C(t)$ is analytic in the entire complex plane \cite{Araki:1969ta}. Accordingly, the slowest possible decay of the power spectrum for large $\omega$ is, schematically, $f^2\propto \omega^{-\omega}$. This corresponds to maximal possible growth of the norm of nested commutators $|[H,[H,\dots A]]|$ allowed by 1D geometry \cite{Avdoshkin_2020}. Corresponding maximal growth of Lanczos coefficients is $b_n \propto n/\ln(n)$ \cite{Parker_2019}.}

For {\it any} field theory and a local operator $A(t)$ (or a local operator integrated over some region),   singularity of the two-point function when the operators collide immediately implies \eqref{fasym} with $\omega_0=2/\beta$. Hence, trivially, asymptotic behavior of $f^2(\omega)$ cannot be used as a probe of chaos. This prompts the question is if perhaps in field theory $b_n$ may exhibit some complex behavior such that asymptotic behavior of $b_n$ may distinguish integrability from chaos, while \eqref{fasym} wold always  be satisfied.  A study of conformal field theories in flat space in \cite{Dymarsky:2021bjq} revealed that nothing of the sort happens and $b_n$ demonstrate a ``vanilla'' asymptotic behavior fixed by the singularity of $C(t)$ at $t=i\beta/2$, 
\bea
\label{asympt}
b_n= {\pi\over \beta}(n+\Delta+1/2)+o(n),\qquad n\gg 1, 
\eea
where $\Delta$ is the dimension of the operator $A$ controlling the pole singularity at $t=i\beta/2$, 
\bea
\label{singularity}
C(t)\sim {1\over (t-i\beta/2)^{2\Delta}}.
\eea
Now we ask if this is perhaps a feature of conformal models in flat space that $b_n$  always grow as $b_n\propto n$, while for more complicated QFTs  Lanczos coefficients may exhibit more complex behavior. And indeed, as we see below,  once scale is introduced through a mass or space volume,  coefficients $b_n$ may split into even and odd branches, with different asymptotic behavior, while \eqref{fasym} with $\omega_0=2/\beta$ is always satisfied. 

\section{Free massive fields in flat space}
\label{sec:freemassive}
In this section we consider free massive scalar\footnote{Also see \cite{Camargo:2022rnt}, which similarly considered free massive scalar field and found ``persistent staggering'' $c_o\neq c_e$. The original version of \cite{Camargo:2022rnt} also reported   different values of slopes for odd and even branches of $b_n$, both smaller than $\pi/\beta$. We regard this as a numerical 
artifact -- we show in Appendix \ref{app:a1a2} that in case of two different slopes,  in full generality one slope is always larger than, and the other is always smaller than $\pi/\beta$, see \eqref{alphaineq}.} and Dirac  fermion in $d$ spacetime dimensions, for which the correlation functions are given by,
see Appendix \ref{app:freetheories},
\bea
\nonumber
C(t)={\rm Tr}(e^{-\beta H/2}\phi(t,x) e^{-\beta H/2} \phi(0,x))\propto \int_{\tilde{m}}^\infty (y^2-\tilde{m}^2)^{d-3\over 2} {\cosh (y\, \tau)\over \sinh(y/2)} dy
\eea
and 
\bea
\nonumber
C(t)={\rm Tr}(e^{-\beta H/2}\psi^\dagger (t,x) e^{-\beta H/2} \psi(0,x))\propto 
\int_{\tilde m}^\infty dy \, ( y^2 - \tilde m^2)^{d-3\over 2}  
 {y\cosh(y\tau) \pm \tilde m \sinh(y \tau)\over \cosh\big({y/2}\big)},
%\\ \label{fermion}
\eea
where $\tau \equiv  i t/\beta$, and $\tilde{m}=\beta m$.
In the first case Lanczos coefficients split into even and odd branches, both growing linearly with $n$ albeit with different intercepts, as shown for $d=4$  in the left panel of  Fig.~\ref{bnfreemodels}. We call this behavior  ``persistent staggering'':  $b_n$ grow linearly, with  the universal slope $\pi n/\beta$, but even and odd branches have different finite terms:
\bea
\label{asymptm}
b_n=\left\{\begin{array}{c}
{\pi\over \beta}(n+c_e)+o(n),\qquad {\rm even}\, n.\\
{\pi\over \beta}(n+c_o)+o(n),\qquad {\rm odd}\, n,
\end{array} \right.\qquad n\gg 1. 
\eea 
In the second case of free massless fermions $C(t)$ is not an even function, hence besides $b_n$, Lanczos coefficients also include $a_n$, see Appendix  \ref{app:abn}.
In this particular case coefficients $a_n=\pm (-1)^n \tilde m$ can be expressed analytically, while $b_n$ exhibit standard ``vanilla'' asymptotic \eqref{asympt} $\beta b_n=\pi n$, see the right panel of  Fig.~\ref{bnfreemodels}.

\begin{figure}
\label{bnfreemodels}
\includegraphics[width=0.5\textwidth]{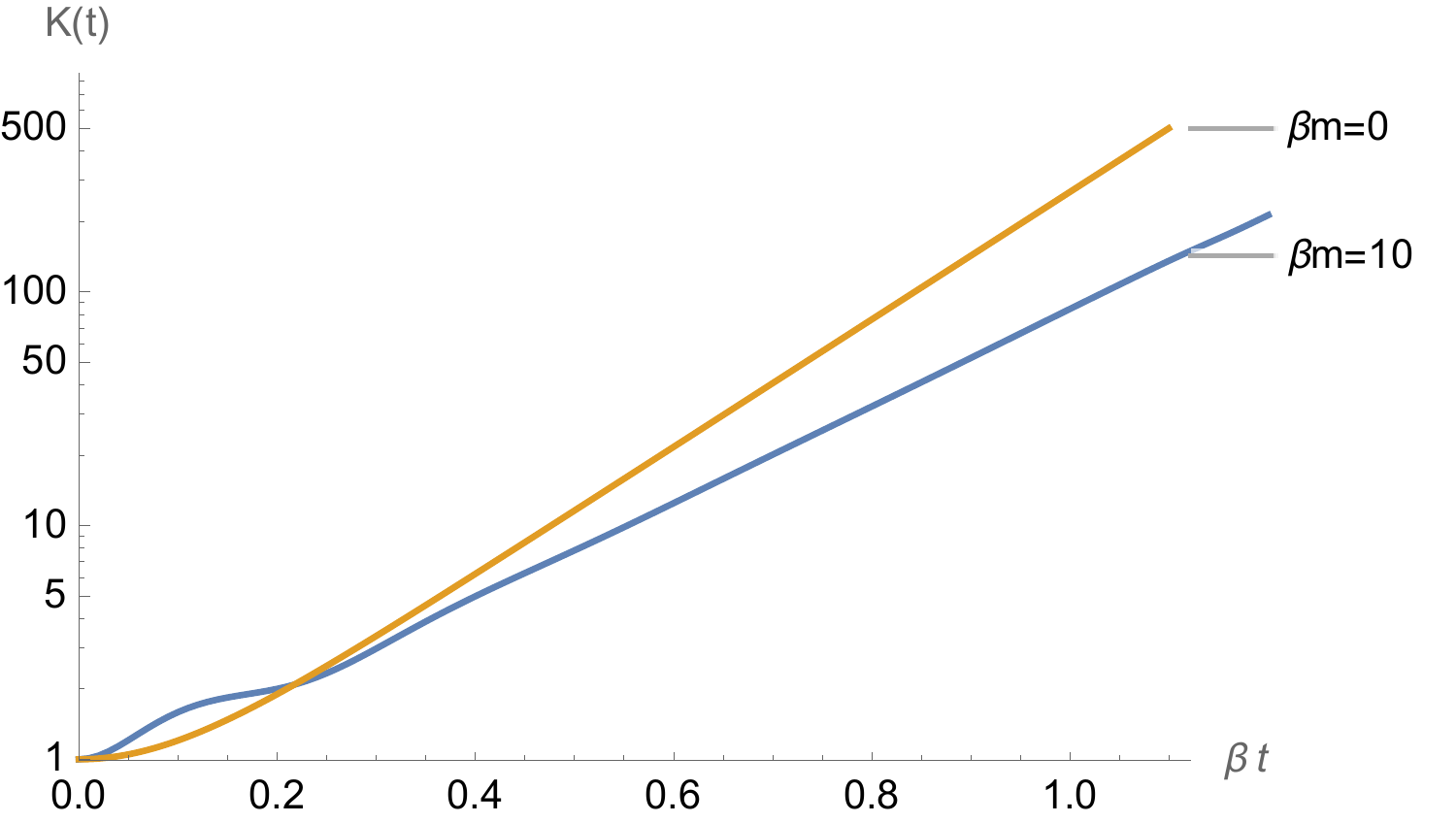}\, \,
\includegraphics[width=0.5\textwidth]{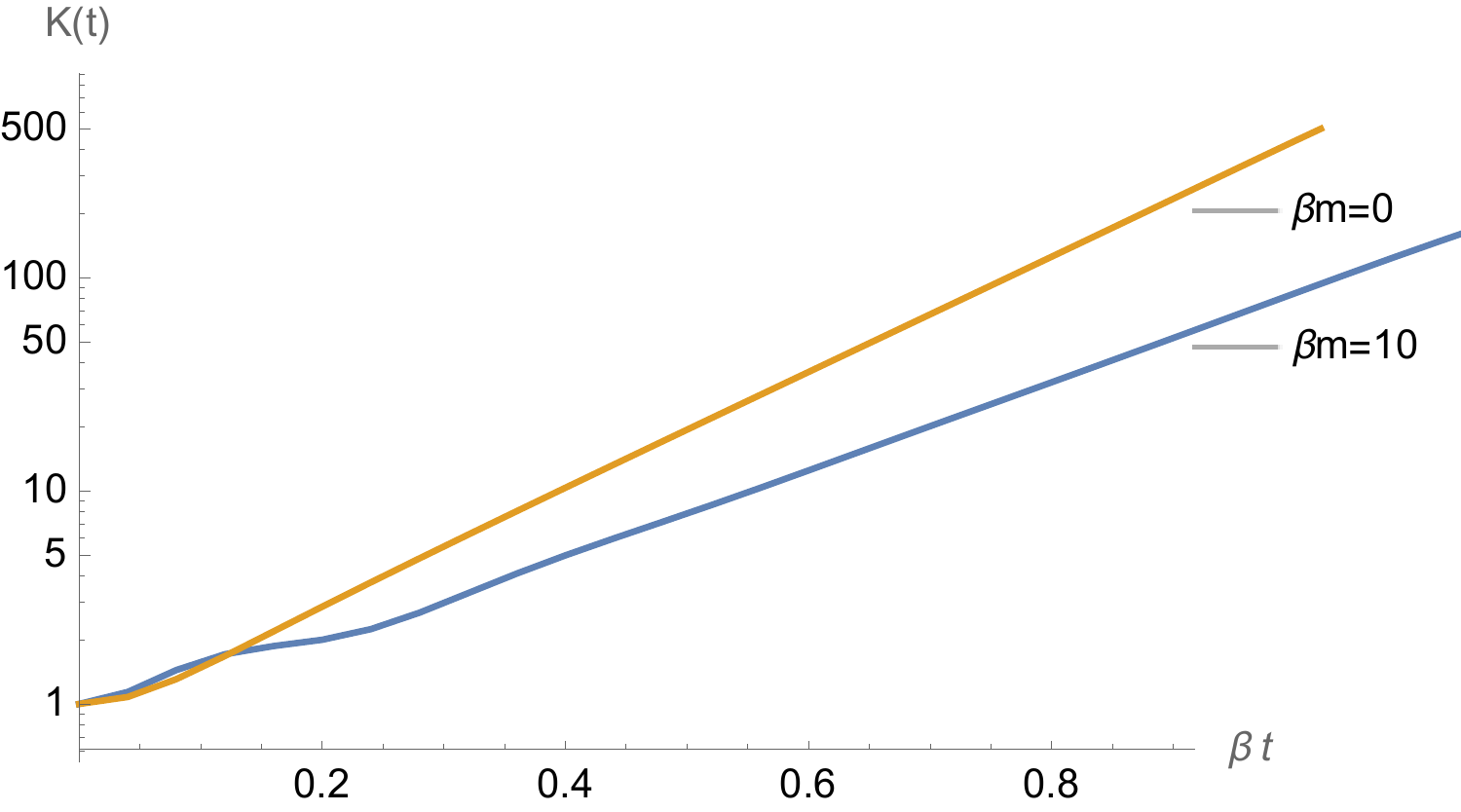}
\caption{Krylov complexity for the free massive scalar (left) and fermion (right) in $d=4$ dimensions. For the massless scalar $K=1+2\sinh^2(\pi t/\beta)$ is known analytically. The main effect of mass is the decrease of Krylov exponent $\lambda_K$. }
\label{Kfree}
\end{figure}

Both the persistent staggering and oscillating $a_n$ have similar effects on Krylov complexity, which we calculate in both cases numerically, see Fig.~\ref{Kfree}. Namely, $\ln K(t)$ continue to grow linearly, but with the slope $\lambda_K$ smaller than $2\pi/\beta$.  This provides a non-trivial test of the second inequality of \eqref{MSS}. The first inequality is satisfied trivially because we assign $\lambda_{\rm OTOC}=0$ in free field theories. 

From the mathematical point of view, it is not entirely clear which property of $C(t)$ is a cause for persistent staggering.  (We are focusing on the case of even $C(t)$ here, when $a_n\equiv 0$.) Based on a handful of examples reference \cite{viswanath2008recursion} claims, apparently erroneously, this is due to periodicity of $C(t)$. We propose that persistent staggering is a reflection of the ``mass gap'' of $f^2(\omega)$, that it is zero for $|\omega|$ smaller than some finite value $\omega_m$.  Proving this statement or ruling it wrong, is an interesting mathematical question. In the Appendix \ref{PSE} we argue that staggering, i.e.~the difference between $c_e$ and $c_o$ in \eqref{asymptm}, can not be deduced from the structure of the singularity of $C(t)$ on the complex plane, unlike the mean value $(c_e+c_o)/2$, which is directly related to  the pole structure of $C(t)$ at $t=i\beta/2$, i.e.~the dimension of the operator $\phi$.  

For free fields non-zero particle mass automatically translates into ``mass gap'' at the level of $f^2$. When  particles are massive but interact, $f^2(\omega \rightarrow 0)$ will not be zero, although will be  exponentially small.\footnote{We thank Luca Delacretaz for brining  this point to our attention.} It is an interesting question to understand if this  behavior will have any obvious imprint on $b_n$, and consequently on Krylov complexity. 

In the case of persistent staggering exponential growth of  $K(t)$ was verified numerically. We leave it as another open question to develop an analytic approximation to evaluate $\lambda_K$ in terms of $\beta, c_1, c_2$. 

\begin{figure}
\label{bnfreemodels}
\includegraphics[width=0.5\textwidth]{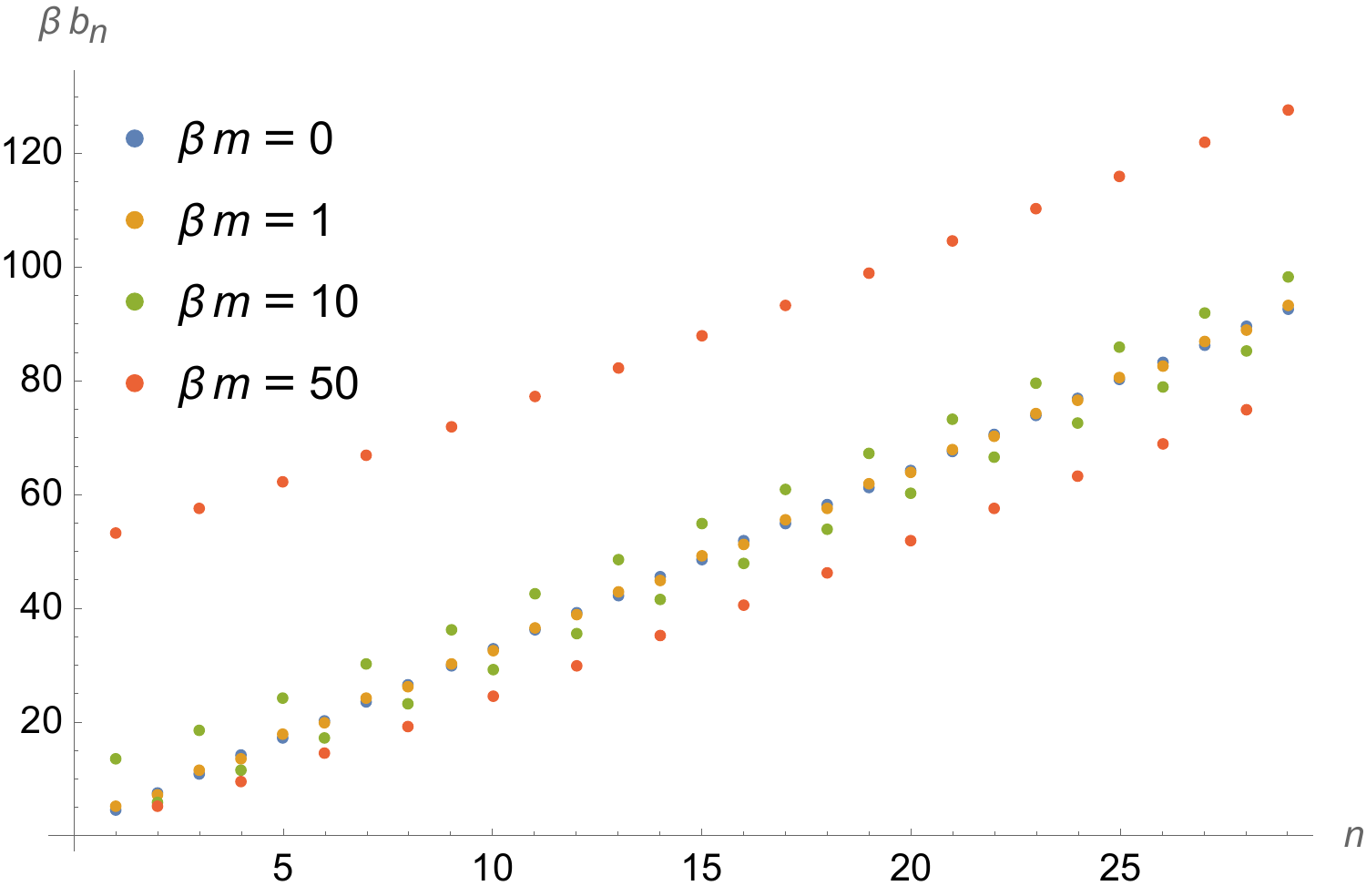}\, \,
\includegraphics[width=0.5\textwidth]{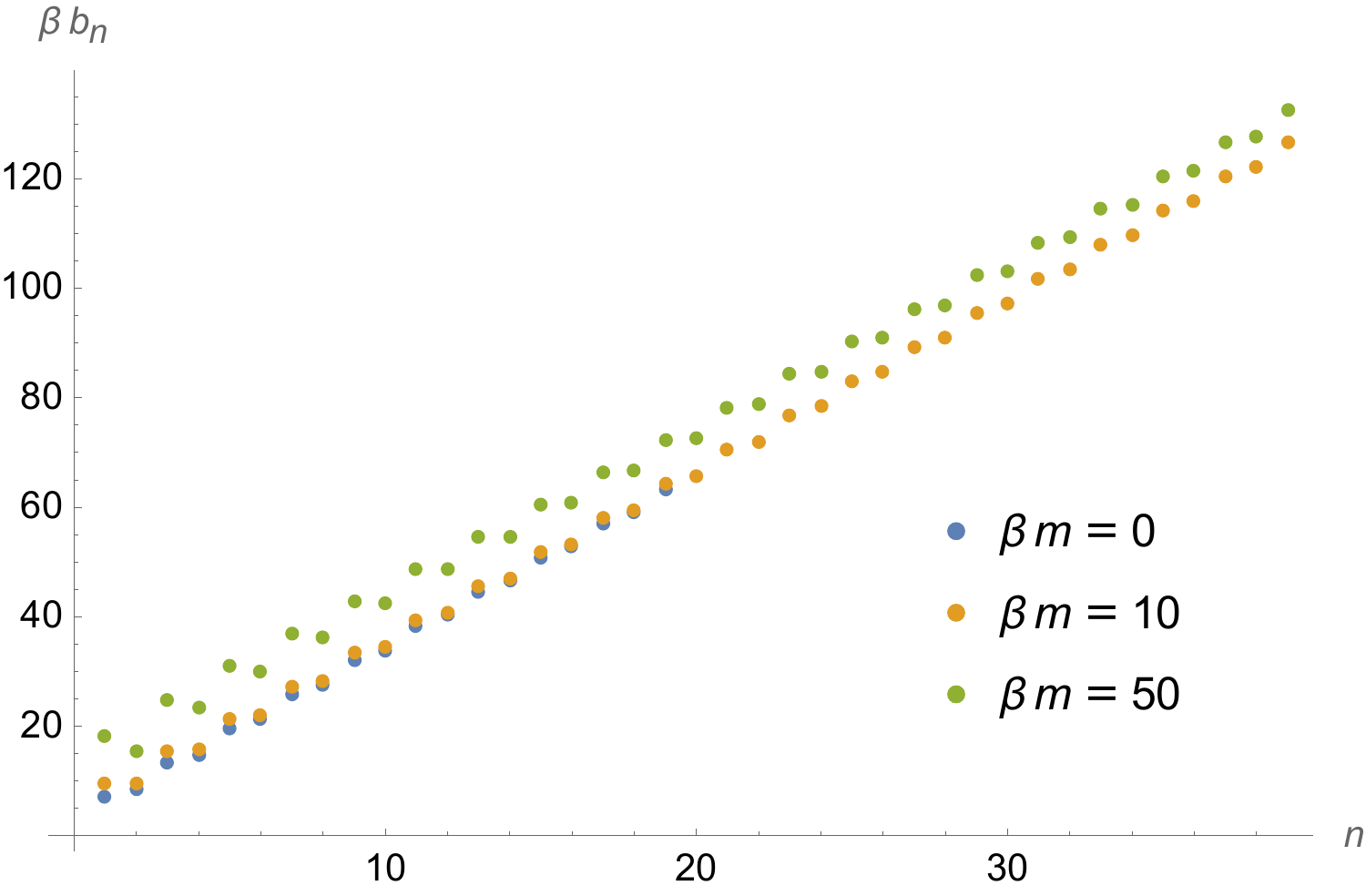}
\caption{Lanczos coefficients for free massive scalar (left) and fermion (right) in $d=4$ dimensions. Values for the  conformal (massless) cases we discussed in \cite{Dymarsky:2021bjq}. For the scalar, introduction of  $\tilde{m}=\beta m$ results in ``persistent staggering'' of $b_n$ around  the conformal values (larger $\tilde{m}$ causes larger staggering amplitude) but does not change the slope $\pi n/\beta$. Lanczos coefficients split into two branches,  see eq.~\ref{asymptm}. For fermions, asymptotically, $b_n$ grow linearly with the slope $\pi n /\beta$ and the $m$-dependent intercept. The main effect of mass is non-vanishing $a_n=\tilde{m}(-1)^n$. }
\label{bnfree}
\end{figure}

\section{CFTs on a sphere }
\label{sec:onS}
In this section we consider a CFT places on a sphere and calculate Lanczos coefficients and Krylov complexity associated with the Wightman-ordered thermal two-point function. We consider two cases,  4d free massless scalar on $\Sp^3$ and holographic theories. 

\subsection{Free scalar on $\Sp^3\times \Sp^1$}
\label{sec:S3S1}
We first consider a  4d free massless scalar compacted on a three-sphere. Because we are considering finite-temperature correlation function the 4d compactification manifold is $\Sp^3 \times \Sp^1$.  Corresponding thermal two-point function is given by, see Appendix \ref{app:S3S1}, 
\bea
\label{nl}
C(\tau,R)&\propto &\sum_{n,\ell \in \Z} {(\tau+1/2+n)^2-(R\ell)^2\over ((\tau+1/2+n)^2+(R\ell)^2)^2}=\\ \label{l}
&&\sum_{\ell \in \Z} {\pi^2\over 2}\left({1\over \cosh^2(\ell \pi R+i\pi \tau)}+{1\over \cosh^2(\ell \pi R-i\pi \tau)}\right)-{2\pi \over R}=\quad \\ \label{n}
&&\sum_{n\in \Z} {\pi^2\over R^2} {1\over \sinh^2((n+1/2+\tau)\pi/R)}. \label{rep3}
\eea
Here $R$ is the radios of $\Sp^3$ measured in the units of $\beta$, which is  the radius of $\Sp^1$. Euclidean time $\tau$  is defined as  $\tau=i t/\beta$.  As expected the function is periodic  in Euclidean time $\tau \rightarrow \tau+1$, but  it is also periodic   in Lorentzian time under
$t\rightarrow t+2 \beta/R$, which follows e.g.~from \eqref{rep3}.

\begin{figure}
\label{bnfreemodels}
\includegraphics[width=0.5\textwidth]{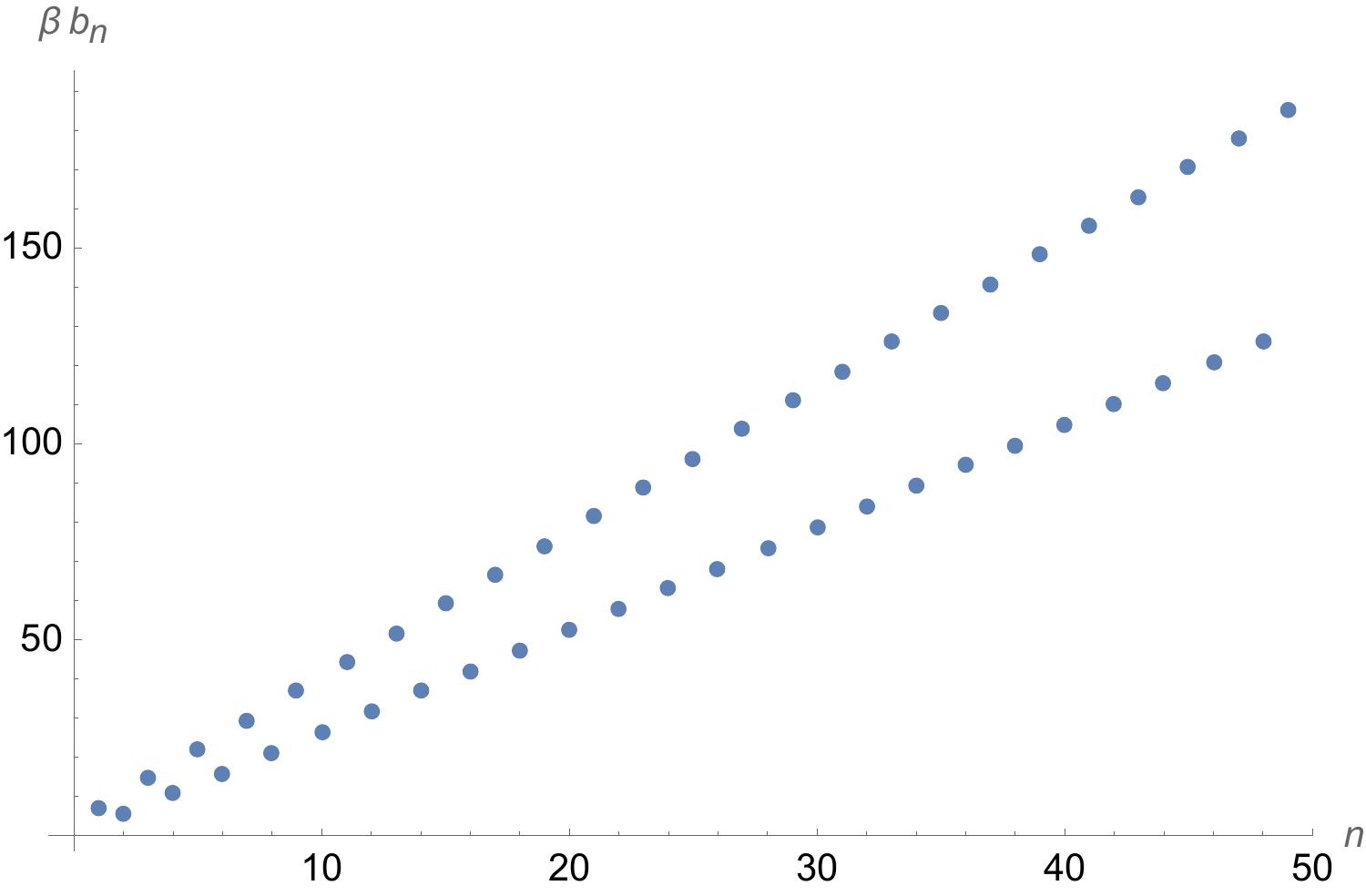}\, \,
\includegraphics[width=0.5\textwidth]{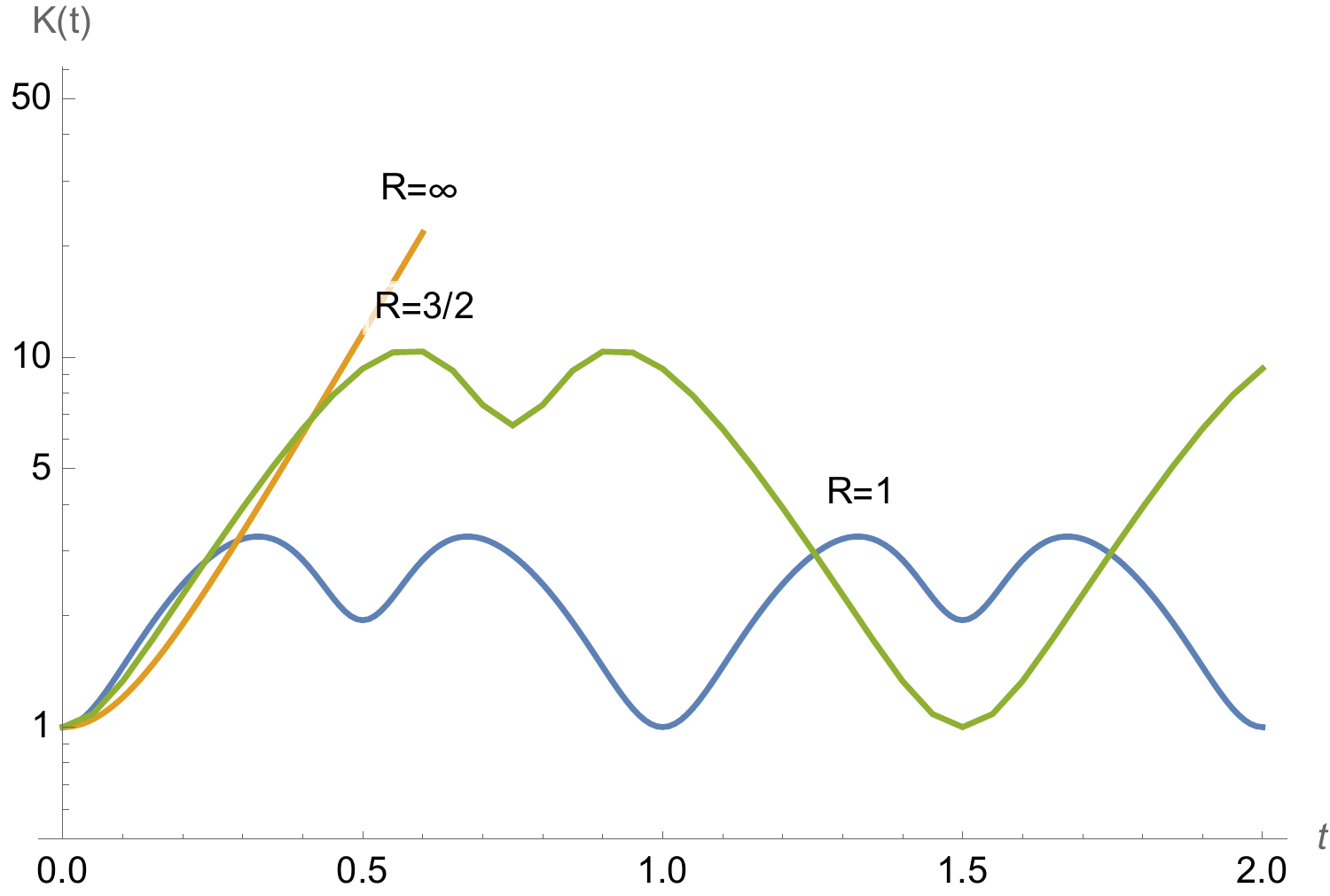}
\caption{Left: Lanczos coefficients for free massless  scalar compactified on  $\Sp^3$ 
of radius $R=1$. Right: Krylov complexity $K(t)$ for radii $R=1$, flat space $K=1+2\sinh^2(\pi t)$, and $R=3/2$.}
\label{S3S1}
\end{figure}

For any finite $R$, Lanczos coefficients split into even and odd branches, which grow linearly with $n$ but with different slopes, see Fig.~\ref{S3S1}. This can be seen analytically in the limit $R\rightarrow 0$ when $C(\tau)$ is exponentially small (unless $|\tau|=1/2$) and Lanczos coefficients, at leading order, are given by
\bea
\label{smallR}
b_n^2=\left({2\pi\over R}\right)^2\left\{ 
\begin{array}{rc}
(n+1)^2/4, &\, \, n=1,3,5,\dots\\
\frac{4 n (n+1)^2}{n+2} e^{-\pi/R}, &\, \, n=2,4,6,\dots
\end{array}
\right.
\eea
Numerically, this approximation is good already for $R\lesssim 1$.  

In the opposite limit  of large $\Sp^3$ the correlation function approaches that one of flat space, plus a correction $-2\pi/R$, plus the exponentially suppressed terms 
\bea
\label{largeR}
C\propto  {\pi^2\over \cos^2(\pi \tau)}-{2\pi \over R}+{\pi^2\over \cosh^2(\pi R-i\pi \tau)}+{\pi^2\over \cosh^2(\pi R+i\pi \tau)}+\dots
\eea
The exponentially suppressed terms are important for the asymptotic behavior of $b_n$. Neglecting them, i.e.~keeping only first two terms in \eqref{largeR} yields the following expression for Lanczos coefficients, which we denote $b^0_n$, 
 \bea
(b_n^0)^2&=&\pi^2 n(n+1)-(-1)^n{c_n^2\over c_n R+d_n},\qquad n\geq 1,\\
c_n&=&\pi  \left(2 n+1-(-1)^n\right),\qquad d_n=4 \left\lfloor \frac{n+1}{2}\right\rfloor \left(\psi ^{(0)}\left(\frac{3}{2}\right)-\psi ^{(0)}\left(\left\lfloor \frac{n+1}{2}\right\rfloor +\frac{1}{2}\right)-2\right). \nonumber
\eea
Here $\psi^0$ is a polygamma function. This expression  has incorrect asymptotic behavior for large $n$. Taking into account two more terms in \eqref{largeR} leads to 
\bea
\label{smalln}
b_n^2=(b_n^0)^2+8 \pi ^2 (-1)^{n} n (n+1) (2 n+1) e^{-2 \pi  R}+O(e^{-\pi R}/R),\qquad n\geq 1,
\eea
which  still has incorrect the asymptotic. We thus conclude that to reproduce ``two slopes'' behavior 
\bea
\label{asymptmS3}
b_n=\left\{\begin{array}{c}
\alpha_{\rm odd}(n+c_1)+o(n),\qquad {\rm odd}\, n.\\
\alpha_{\rm even}(n+c_2)+o(n),\qquad {\rm even}\, n,
\end{array} \right.\qquad n\gg 1. 
\eea 
would require taking into account an infinite serious of ever more exponentially suppressed terms into account.  For all finite $R$, $\alpha_{\rm odd}>\alpha_{\rm even}$; the opposite would contradict $C(t\rightarrow \infty)\rightarrow  0$.\footnote{This follows from existence of a normalized Liuvillian's zero mode if asymptotically,  for odd $n$, $b_n/b_{n+1}<1$.}
The ratio $\alpha_{\rm odd}/\alpha_{\rm even}$ grows with $R^{-1}$, when radius is large two slopes are almost equal to each other and $\pi/\beta$, see section \ref{E4}, but for small $R$ they are significantly different, with $\alpha_{\rm even}$ quickly approaching zero. 
We also show in full generality in the Appendix \ref{E2}, that in case of two-slopes behavior \eqref{asymptmS3}, the inequality $\alpha_{\rm odd}\geq \pi/\beta\geq \alpha_{\rm even}$ always holds, see \eqref{alphaineq}.

Now we discuss  Krylov complexity, which we calculate numerically. We find that $K(t)$ first increases, but then reaches its maximum and starts oscillating quasi-periodically, see the right panel of Fig.~\ref{S3S1}. This behavior is more pronounced for small $R$,  while for large radius the behavior of $K(t)$ initially  follows the flat space counterpart $K(t)=1+2\sinh^2(\pi t)$, before peaking at some finite value. 

The two slopes behavior of $b_n$ is a novel phenomenon, not observed for physical systems  previously. This goes beyond the universal operator growth hypothesis of \cite{Parker_2019}, which assumes a smooth asymptote of $b_n$. Similarly, capped Krylov complexity with the maximal value dependent on $R$ (the ratio of $\Sp^3$ and $\Sp^1$ radii) but not on the UV-cutoff is in tension with the proposal that qualitatively  $K(t)$  is similar to holographic and computational complexities \cite{Barbon:2019wsy,Rabinovici_2022,Rabinovici2022}. In the latter case, for a QFT on compact space,  complexity would grow up to exponentially large values, controlled by the volume of $\Sp^3$ measured in the units of the UV cutoff. In case of $K(t)$ this growth was presumably supposed to come from the region of $b_n$ which is controlled by UV physics (and where $b_n$ are approximately constant while $K(t)$ grows linearly, see sections \ref{sec:XY}, \ref{sec:1D} and Figs.~\ref{fig:XYi} and \ref{fig:1D}). But the example above shows that the operator can be confined at the beginning of ``Krylov chain,'' in which case the  values of $b_n$ for large $n$ do not matter. 

Physically, it is tempting to relate two slopes behavior of $b_n$ and bounded $K(t)$ to finite volume of the space QFT is placed on, but we will see in the next section this is not true in full generality.

Mathematically, the two slopes behavior of $b_n$ is probably due to periodicity of $C(t)$ in Lorentzian time, or more broadly due to $f^2(\omega)$ being a sum of delta-functions for a not necessary periodic grid of values of $\omega$. We leave a proper investigation of this question for the future, together with the question of quantitatively relating $\alpha_1,\alpha_2$ to  properties of $C(t)$. Here we only make one step in this direction and use the integral over Dyck paths developed in \cite{Avdoshkin_2020} to relate a particular combination of $\alpha_1,\alpha_2$ to the location of pole of $C(t)$ in the complex plane, see Appendix \ref{app:a1a2}. Another question, which we also leave for the future, is to analytically relate maximal or time-averaged value of $K(t)$ to $\alpha_i$. 

\subsection{Holographic theories}
Next we consider a holographic theory with the two-point function of heavy operators given by the sum over  geodesic lengths in thermal AdS space, or black hole background, below and above the Hawking-Page transition correspondingly. In the former case the two-point function  is\footnote{We thank Matthew Dodelson for collaboration laying foundation for results of this section.}
\bea
\label{TAdS}
C(t) \propto \sum_n {1\over (\cosh((\tau+1/2+n)\beta)-1)^{\Delta}} 
\eea
where the radius of boundary sphere is taken to be one. This expression is valid for  all$\Delta$, not necessarily large,  see e.g.~\cite{Alday:2020eua}.
This expression is valid in all dimensions $d\geq 3$. We notice that, again, besides periodicity in Euclidean time $\tau\rightarrow \tau+1$, this function is periodic in Lorentzian time $t \rightarrow t+2\pi$. This is presumably the reason why $b_n$ behave qualitatively the same as in the previous subsection -- they split into two branches for even and odd $n$, with both exhibiting asymptotic linear growth \eqref{asymptmS3}. The behavior of Krylov complexity also follows the pattern of free scalar on $\Sp^3$, it first grows but then oscillates quasi-periodically. We thus conclude that in holographic theories Krylov complexity can be UV-independent, thus making it qualitatively different from the holographic complexity \cite{Susskind:2014rva,Brown:2015bva,Brown:2015lvg,Ben-Ami:2016qex,Chapman:2016hwi,Carmi:2017jqz,Belin:2021bga}.

\begin{figure}
\label{bnfreemodels}
\includegraphics[width=0.5\textwidth]{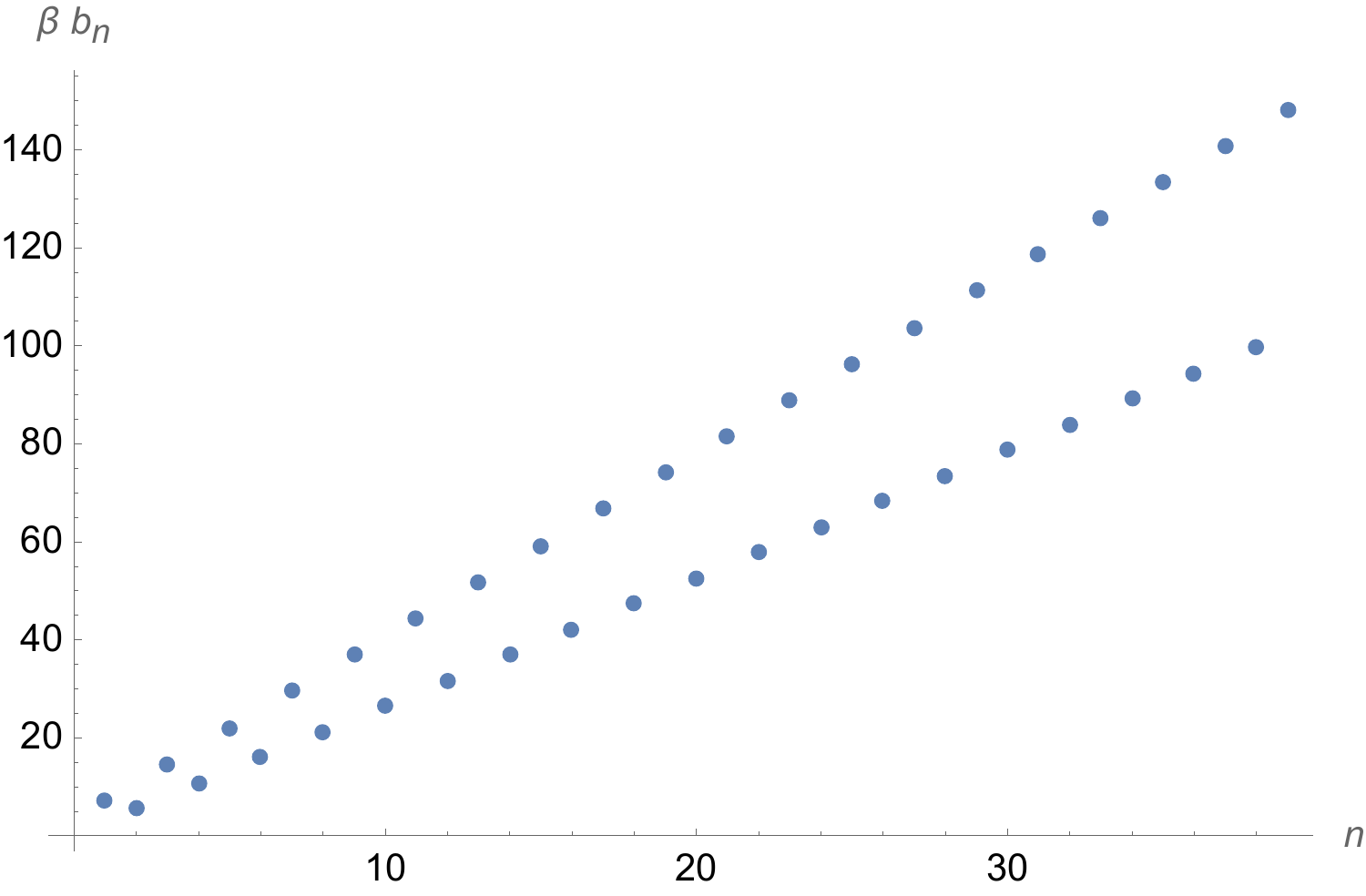}\, \,
\includegraphics[width=0.5\textwidth]{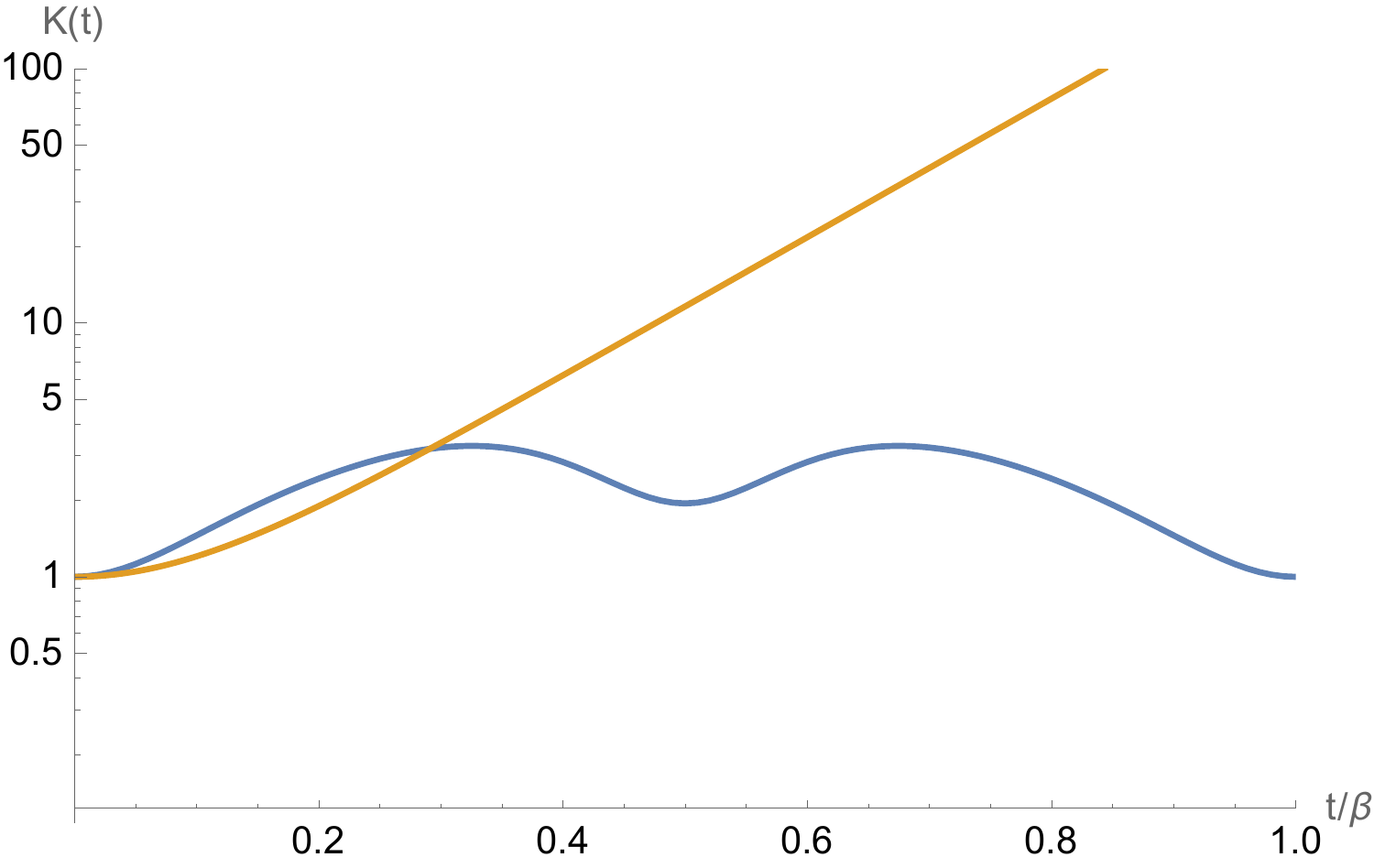}
\caption{Left: Lanczos coefficients for thermal AdS background \eqref{TAdS}  with $\beta=2\pi$ and $\Delta=1$. Right: Krylov complexity for these parameters (blue) superimposed with $K(t)$ for flat space 2d CFT result $K=1+2\sinh^2(\pi t)$ (orange).}
\end{figure}

The discussion above applies to temperatures small enough, below the Hawking-Page transition. As the temperature increases the dual geometry is given by the  BTZ background, assuming we focus specifically on the case of 2d theories. The two-point function in this background is given by \cite{Keski-Vakkuri:1998gmz,Maldacena:2001kr}
\bea
\label{BTZ}
C(t) \propto \sum_n {1\over \left(
\cos(2\pi \tau)+\cosh(4 \pi^2n/\beta)
\right)^{\Delta}}.
\eea
This function is, of course, periodic under $\tau\rightarrow \tau+1$, but there is no periodicity in Lorentzian time. As a result Lanczos coefficients have the ``vanilla'' behavior,  growing linearly with the asymptote \eqref{asympt}. In fact  leading contribution comes from the $n=0$ term in \eqref{BTZ}, which is the same as the flat-space expression, with other terms for $\beta<2\pi$ giving very small corrections. 
Thus, numerically, $b_n^2$ are very close to $\pi^2 (n+1)(n+2\Delta)/\beta^2$  \cite{Dymarsky:2021bjq}.
Accordingly, Krylov complexity grows exponentially with $\lambda_K=2\pi/\beta$.

The calculation in the BTZ background shows that two slopes behavior and capped $K(t)$ are not universal features of QFT on compact spaces, and the unbounded exponential growth of $K(t)$ is possible in the holographic settings. This prompts the question of how the inequality \eqref{MSS} fairs in different holographic scenarios. We see that above the Hawking-Page transition  Krylov exponent $\lambda_K$ is well defined and equal to $2\pi/\beta$. This means second inequality in \eqref{MSS} is saturated and reduces to Maldacena-Shenker-Stanford bound, which is also saturated by holographic theories. Below the transition, Krylov complexity is bounded, hence $\lambda_K$  is not well-defined, or we can formally take it equal to zero. Exactly in the same scenario the OTOC correlator exhibits periodic recurrences, which also makes $\lambda_{\rm OTOC}$ ill-defined (or vanish) \cite{Anous:2019yku}. In other words, \eqref{MSS} remains valid in both cases.

\section{Temperature and UV-cutoff dependence of Lanczos\\  coefficients}
\label{sec:TandUV}
We have seen in the previous sections that giving particles mass or placing a CFT on a compact background both have an explicit imprint on Lanczos coefficients and Krylov complexity. One feature, which nevertheless seems to be lost in the QFT case is the sensitivity of $b_n$ to chaos. Indeed, in all cases above $b_n$ exhibit linear growth (albeit sometimes by splitting into two branches), even though many considered theories are not chaotic. This mirrors the behavior we previously observed for CFTs in flat space \cite{Dymarsky:2021bjq} and complements recent observation that  non-chaotic systems with saddle-dominated scrambling  also exhibit linear growth of $b_n$ \cite{Bhattacharjee_2022}.

This could be our final conclusion -- that asymptotic of Lanczos coefficients is not a proper probe of chaos, but apparent success of this approach in case of discrete systems with finite local Hilbert space, in particular spin chains, hints this conclusion could be premature. 
 As we mentioned above, for 1D  lattice systems slowest possible decay of $f^2(\omega)$ is not exponential but super-exponential  $f^2(\omega)\sim \omega^{-\omega}$ \cite{Avdoshkin_2020}. Yet for integrable spin chains $f^2(\omega)$ decays faster, as a Gaussian (which correspond to asymptotic growth $b_n \propto n^{1/2}$) or even becomes zero for $\omega$ exceeding certain size-independent threshold. This is is the case in all known examples,  including preliminary numerical evidence for the XXZ model \cite{PhysRevE.100.062134}.  On the contrary for the non-integrable 1d Ising model in traverse field it was recently proved in \cite{Cao_2021}  that $f^2$ will decay with the slowest possible asymptotic, $f^2(\omega)\propto \omega^{-\omega}$.
 These results constitute non-trivial evidence for  the ``asymptotic of $b_n$ as a probe of chaos'' stronger version of the universal operator growth hypothesis.\footnote{The original hypothesis claimed that a generic system would feature maximal possible growth of $b_n$.} We should mention, there is an additional evidence for the this hypothesis for D>1 lattice models: an analytic proof of $f^2\propto e^{-\omega/\omega_0}$ asymptotic for a non-integrable 2D lattice model \cite{bouch2015expected}, and for the limit of SYK model \cite{Parker_2019}.

Apparent success with the spin chains and other discrete models and failure in field theoretic models suggest the issue could be related to the continuous nature of the latter. We elaborate on this in the next section.

\subsection{Asymptotic of $f^2(\omega)$ and locality}
While for the spin chains high frequency asymptotic of $f^2$ apparently contains some dynamical information, for continuous systems it is completely universal. 
Let's consider  the Fourier of the two-point correlation function in field theory. Then $f^2(\omega)$  has a simple interpretation as the sum of  transition amplitudes between the one-particle states with energies $E_i$ and energies $E_f=E_i+\omega$. Its high-frequency asymptotic can be deduced from the following consideration.
Let us consider a free quantum-mechanical particle propagating in 1D. Assuming the particle is fully localized, Heisenberg uncertainly principle implies the transition amplitude to a state with arbitrarily high energy is not suppressed (the Fourier of delta-function is a constant). After restoring $\beta$-dependent factors this means $f^2(\omega)$ will be proportional to $e^{-\beta \omega/2}$. Here locality of the initial state was crucial for the transition amplitude to states with large final energy to be unsuppressed. Going back to the two-point function $C(t)$, this translates into locality of the operator $A$. 
To see this in more detail, let us consider a same example, free quantum-mechanical particle with the Hamiltonian $H={p^2\over 2m}$, and an operator in the Heisenberg picture
\bea
A=e^{-x^2/a}.
\eea 
We would like to evaluate \eqref{C}
%\bea
%C(t)={\rm Tr}(e^{-\beta H/2}A(t) e^{-\beta H/2} A(0))=\int d\omega f^2(\omega) e^{-i \omega t},
%\eea
and find
\bea
f^2(\omega)=K_0(\beta(\beta+4am)\omega/2).
\eea
It's clear that when $a$ is large, which corresponds to the de-localized wave-packet, the decay of $f^2$ is {\it faster} then the kenematicaly fixed value $f^2\propto e^{-\beta \omega/2}$ associated with the point-like operator $a\rightarrow 0$. 

Provided the logic above is correct we may expect that in the case of a translationally-invariant lattice models similar behavior will emerge in the  continuous (long wave) approximation, emerging in the low energy limit. We test this below in case of  bosonic and fermionic lattice models. For the fixed value of coupling constants, effective wavelength is specified by inverse temperature $\beta$. Thus we expect that for large $\beta$ an approximate continuous description will emerge, leading to universal high frequency tail of $f$ and as well as universality of $b_n$ behavior. On the contrary, for small $\beta$ we expect to recover the diversity of $b_n$ behaviors previously observed in the literature, and in particular connection between the asymptotic of $b_n$ and chaos.

\subsection{Integrable $XY$ model at finite temperature}
\label{sec:XY}
\begin{figure}
\label{bnfreemodels}
\includegraphics[width=0.5\textwidth]{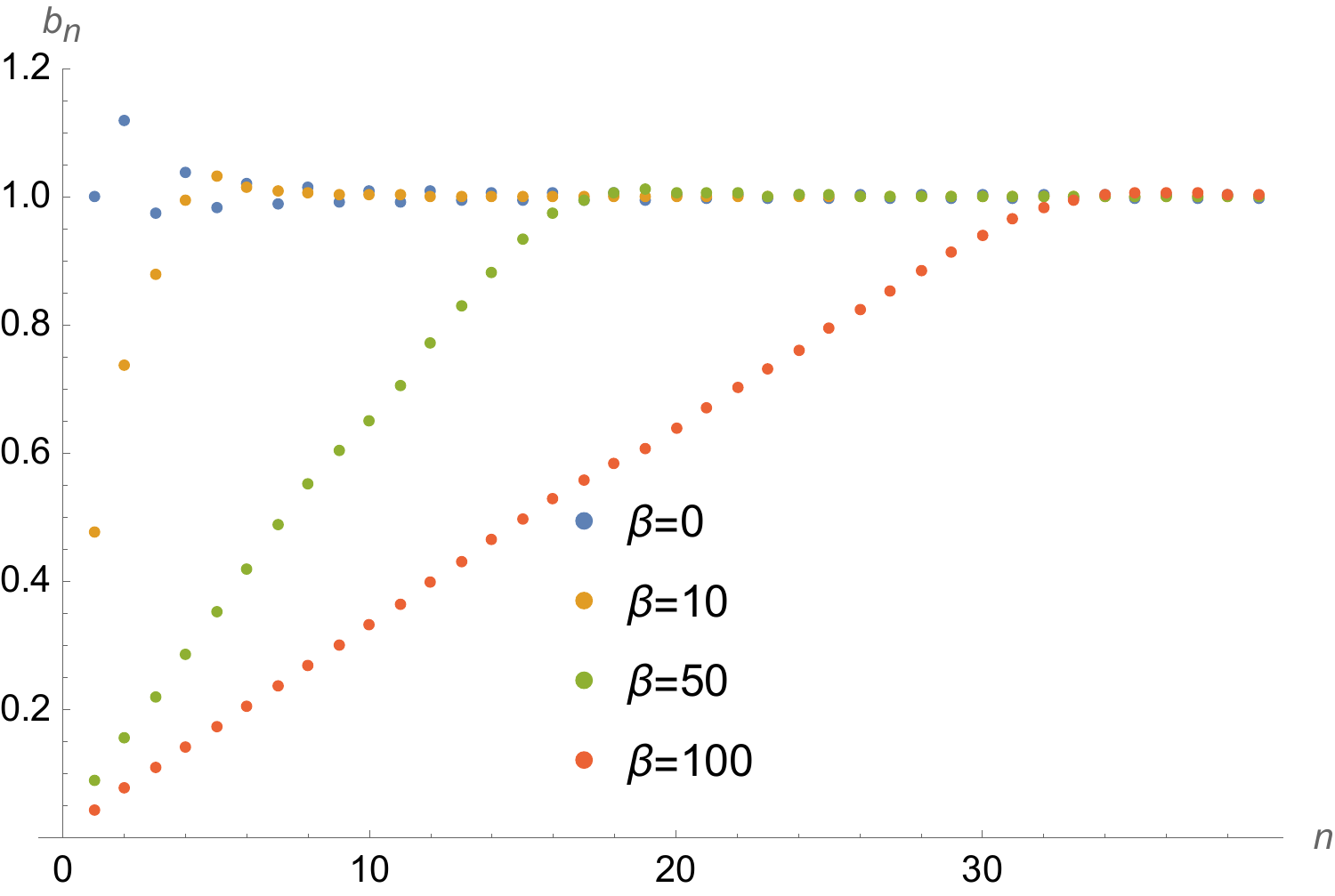}\, \,
\includegraphics[width=0.5\textwidth]{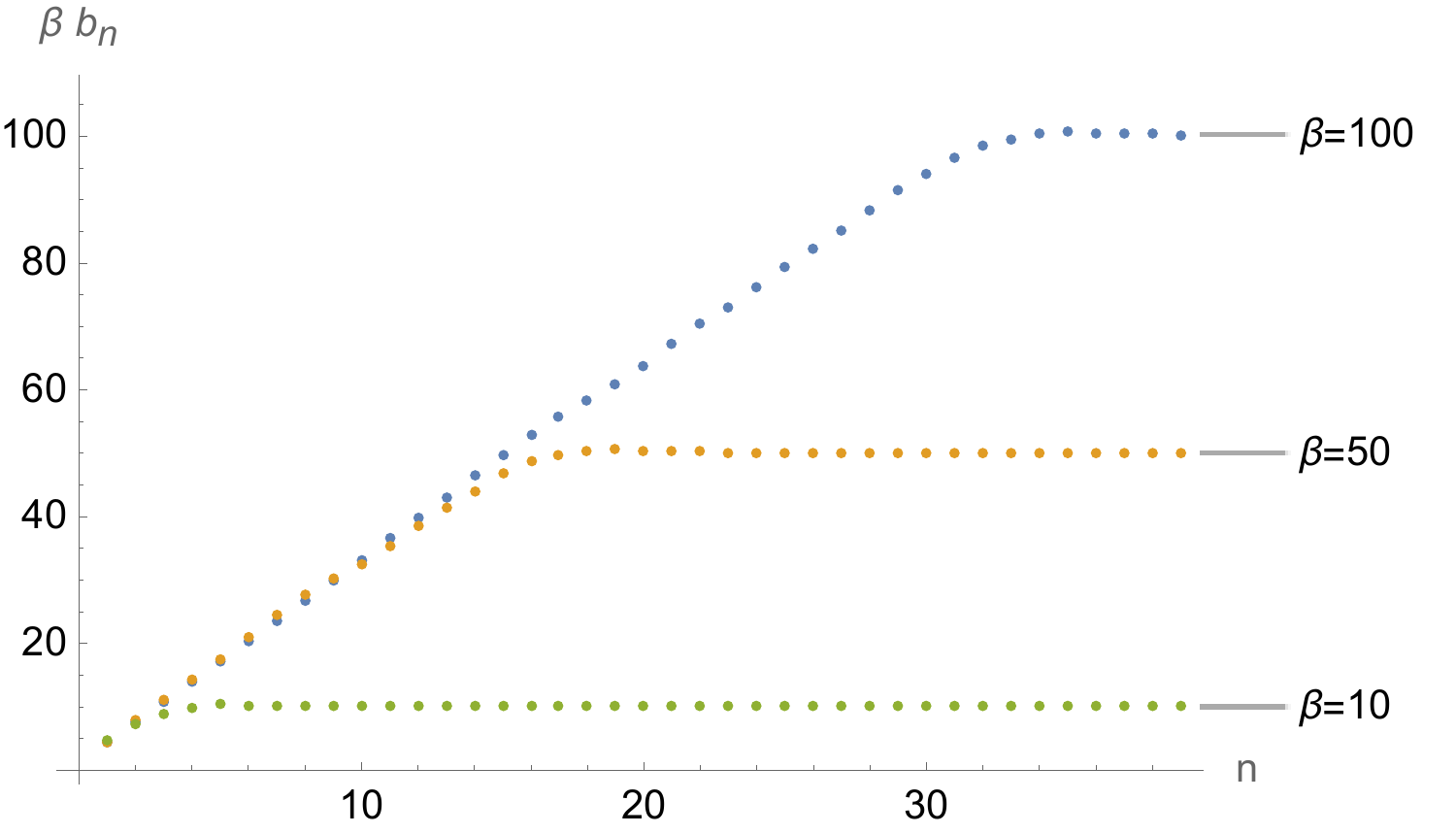}
\caption{Both panels: Lanczos coefficients for the isotropic XY model \eqref{XYi}. }
\label{fig:XYi}
\end{figure}

The integrable periodic $XY$ spin chain 
\bea
H=\sum_{i=1}^N (1+\gamma) S_i^x S_{i+1}^x+(1-\gamma)S_i^y S_{i+1}^y  - h\, S_i^z
\eea
is solvable through the Jordan-Wigner transform and the two-point function  at finite temperature 
\bea
C(t)=\langle S^z (t) S^z(0)\rangle^W_\beta={\rm Tr} (e^{-\beta H/2}S^z (t) e^{-\beta H/2}S^z)
\eea
is known explicitly, see Appendix \ref{app:spinchain}. We are interested in the thermodynamic limit of infinite $N$ and  start with the case of isotropic XY model with $\gamma=h=0$, in which case 
\bea
\label{XYi}
C(\tau)={1\over 4\pi^2}\left(\int_{-1}^1 {dc\over \sqrt{1-c^2}}{\cosh(c\, \tau)\over \cosh(c\, \beta/2)}\right)^2.
\eea
Calculating $b_n$ numerically, for different values of $\beta$ reveals the following behavior. Initially Lanczos coefficients grow linearly as $b_n \approx \pi n /\beta$, and then saturate at some universal value $b_n\approx 1$, see Fig.~\ref{fig:XYi}. Clearly, initial region of linear growth increases for larger $\beta$. This has the following interpretation, at small temperatures the isotropic XY model becomes the free fermion CFT, hence Lanczos coefficients exhibit  universal vanilla CFT behavior \eqref{asympt}. For large $n$, when $b_n$ (which have dimension of energy) become of the order of UV-cutoff (which is of order one in our case), the universal QFT behavior is substituted by the true asymptotic behavior reflecting the dynamics of the lattice model. Since the isotropic XY model is free, Lanczos coefficients saturate at a fixed value, in full agreement with the previous observations that the asymptotic of $b_n$ is a probe of chaos (or lack thereof). 

The main takeaway here is that to probe chaos, one should consider true asymptotic behavior of $b_n$ which is controlled by UV-cutoff physics. In the QFT models with an infinite cutoff, such as conformal field theories, the true asymptote is not accessible, giving way to the essentially universal vanilla behavior (of one or two linearly growing branches of $b_n$).

\subsection{Free bosons on the lattice}
\label{sec:1D}
Next we consider  free oscillators on the 1D lattice with periodic boundary conditions,\footnote{Scalar field with UV-cutoff was also considered in \cite{Camargo:2022rnt}.} 
\bea
H={1\over 4} \sum_{i=1}^N \pi_i^2+(\phi_{i+1}-\phi_i)^2+4\mu^2 \phi_i^2,
\label{1Dboson}
\eea
which becomes 1D free  massive scalar theory in the continuous limit. We can think of this model as the 1+1 free scalar QFT with a cutoff. In the thermodynamic limit $N\rightarrow \infty$, when the mass $\mu$ and temperature $\beta^{-1}$ are chosen to be much smaller than the cutoff, $\beta\gg 1 \gg \mu$, we expect the CFT in flat space behavior with $f^2\propto e^{-\beta \omega/2}$ and linearly growing Lanczos coefficients $b_n \sim \pi n/\beta$. At high energies we are dealing with the discrete integrable model of non-interacting particles with the energies belonging to a finite width zone. Accordingly $f^2(\omega)$ vanishes for $\omega$ exceeding certain value $\omega^*$. Thus, similarly to the isotropic XY model considered above, Lanczos coefficients are expected to approach a constant, signaling lack of chaos.  

\begin{figure}
\includegraphics[width=0.49\textwidth]{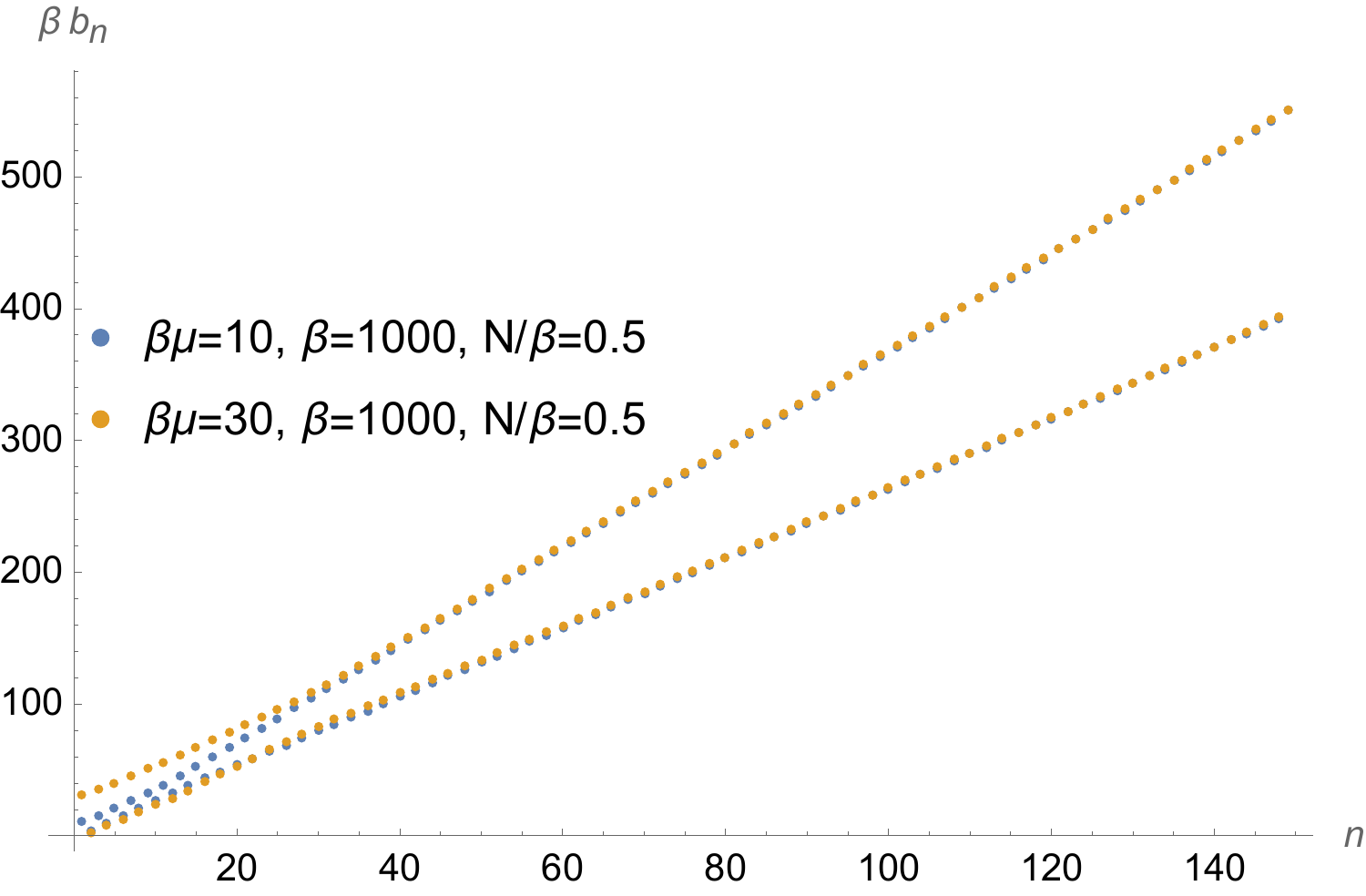}\, \,
\includegraphics[width=0.49\textwidth]{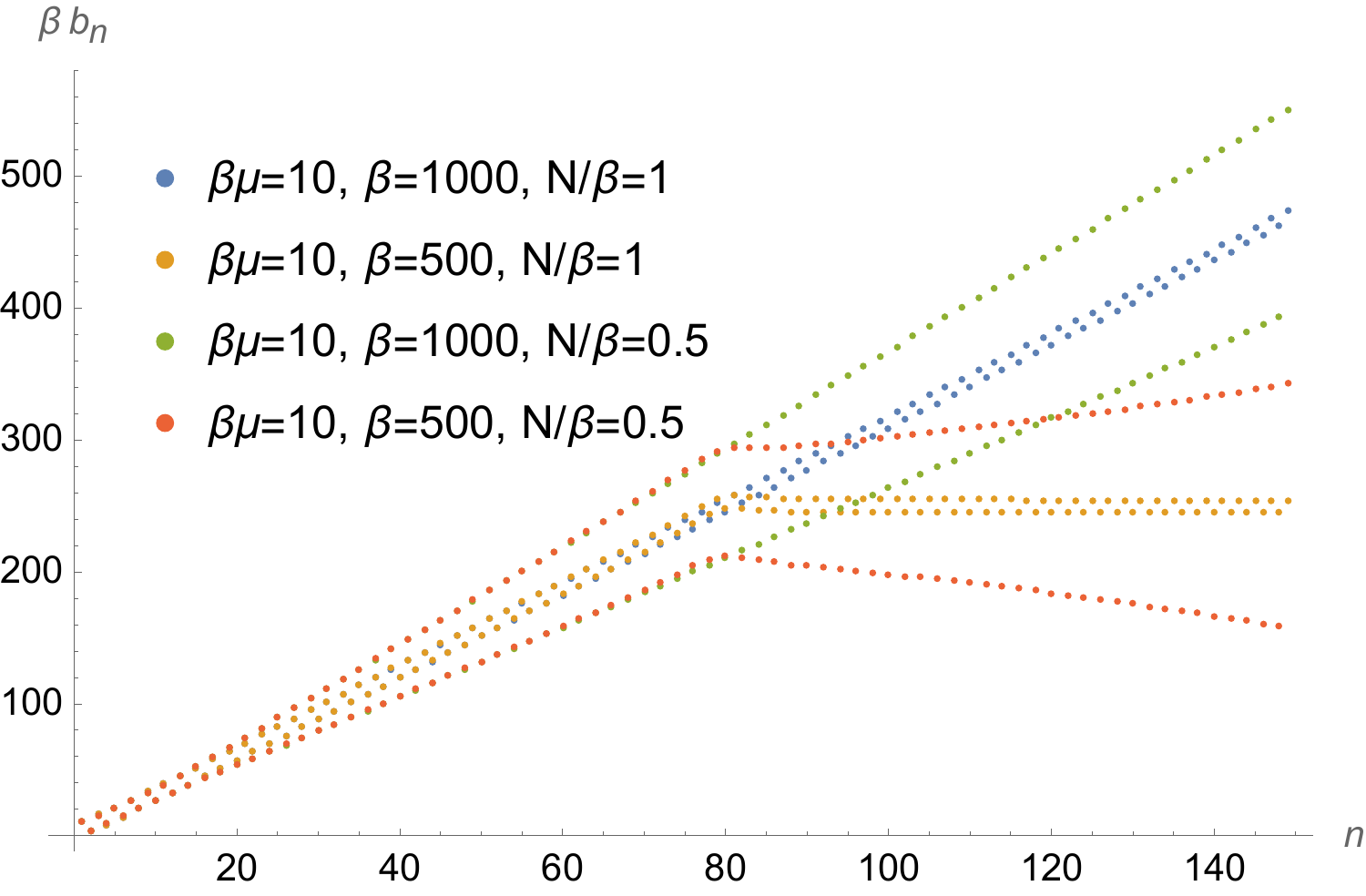} 
\caption{Lanczos coefficients for free bosons on the lattice \eqref{1Dboson} for different values of $\beta,\mu,N$ parameters. For large $\beta$ initial behavior of $b_n$ is that one in filed theory -- coefficients grow linearly, either demonstrating persistent staggering and/or two-slopes behavior, see left panel. When $b_n$ approach UV-cutoff value the QFT behavior changes into one governed by the UV physics, see right panel. }
\label{fig:1D}
\end{figure}

If  $N$ is finite,  in the long-wave limit the system is is described by free massive  QFT placed on $\Sp^1$. In other words, the model \eqref{1Dboson} includes all three deformations we discuss in the paper: the mass, compact spatial manifold and the UV-cutoff. Accordingly, depending on the interplay between $\beta, \mu$ and $N$, at the level of $b_n$ we expect to see the combinations of some or all three features: persistent staggering, two-slopes behavior (since the theory is free,  and the spectrum of excitations is equally-spaced), and field theory behavior below and saturation of $b_n$ near the UV-cutoff.   An explicit calculation of \eqref{C} yields
\bea
C(\tau) \propto {1\over N} \sum_{{\rm k}=1}^N {1\over \epsilon_k} {\cosh (\epsilon_k \tau)\over \sinh(\epsilon_k \beta/2)},\qquad \epsilon_k=\sqrt{\sin^2(k)+\mu^2},\quad k={\pi {\rm k}/N}.
\eea
and the behavior of Lanczos coefficients for different $\beta$, $N$ and $\mu\ll 1$ is shown in Figure~\ref{fig:1D}.

As expected, for essentially infinite  $N$ and large $\beta$ initially $b_n$ grow linearly, as $b_n\approx \pi n/\beta$, but once the value of $b_n$ becomes the order of the cutoff (which is of order one in our case), they saturate to a constant. The punchline here is clear: for small temperatures the prolonged universal linear growth of $b_n$, dictated by the locality and Heisenberg uncertainty principle, will eventually give way to true  asymptotic behavior governed by the discrete model emerging at the UV-cutoff scale.  When $N$ is finite we observe the two slopes behavior, on top of the persistent staggering due to mass, and the  transition toward true asymptote due to UV-cutoff.\footnote{In this case $C(t)$ is not periodic in Lorentzian time, but for any finite $N$, $f^2(\omega)$ has finite support, which is perhaps behind the two slopes behavior,  clearly visible in Fig.~\ref{fig:1D}.} 
%
%Going back to local field theory, i.e. without any explicit cutoff, we can think of it as the low energy limit of some lattice model with an arbitrarily larger $\Lambda_{\rm UV}$. Hence the linear growth of $b_n$ in field theory should not be understood as an asymptotic behavior, but only intermediate behavior, which is fixed kinematicaly, while true asymptotic behavior potentially sensitive to integrability or chaos will emerge for $b_n$ of order $\Lambda_{\rm UV}$. 
%
%When the cutoff is finite, the size of linear growth region depends on $\beta$. It is approximately of the size $\beta \Lambda_{\rm UV}$, excluding first few $n$ before the linear growth sets it. Further evidence behind this picture  is provided by an integrable spin chain at finite temperature, considered in the next section.

\section{Discussion}
\label{sec:discussion}
In the paper we calculated Lanczos coefficients $b_n$ and Krylov complexity $K(t)$ of local operators in several models of quantum field theory, free massive scalars and fermions, massless scalars compactified on a sphere, and a few holographic examples. Our calculations revealed that all three deformations of CFTs in flat space: giving particle mass, placing theory on a compact space, and introducing finite UV-cutoff has a clear imprint on $b_n$ and $K(t)$. Namely, mass leads to persistent staggering of $b_n$ \eqref{asymptm} and decreases the Krylov exponent $\lambda_K$, while $K(t)$ still grows exponentially. Compact space, at least in certain cases, leads to two sloped behavior \eqref{asymptmS3} and a capped Krylov complexity. Finite cutoff introduces a  new  asymptotic regime, where $b_n$ behavior is controlled by the lattice model at the UV scale. 

These examples help us formulate several takeaway messages, clarifying the universal operator growth hypothesis of \cite{Parker_2019} and the role of Krylov complexity. First, asymptotic behavior of $b_n$ in  physical systems goes beyond universality previously discussed in the literature, namely $b_n$ can split into two branches for even and odd $n$, each with its own asymptotic behavior. In particular this means that 
asymptotic behavior of $b_n$ is not mathematically equivalent or fully controlled by the high frequency asymptote of $f^2(\omega)$. 
Second, we clarified the role of temperature for $b_n$ in lattice models. 
Third, in field theory with an infinite UV-cutoff Lanczos coefficients do not probe chaotic behavior. But when a finite cutoff $\Lambda_{\rm UV}$ is introduced, by embedding the QFT  into a lattice model, 
 a true asymptotic regime of $b_n\gtrsim  \Lambda_{\rm UV}$ emerges, which is conjecturally probing chaos in the underlying lattice model, modulo issues raised in \cite{Bhattacharjee_2022}.  Here it is worth noting that a chaotic QFT can not emerge as a long wave limit from an integrable lattice model. Thus, with help  of a finite UV-cutoff, universal operator growth hypothesis can be extended to quantum field theories.  

Another takeaway message is the generalization of the Maldacena-Shenker-Stanford bound \eqref{MSS}, which we tested in several non-trivial settings. 

Finally,  the examples of  theories demonstrating two slopes behavior of $b_n$ and a capped $K(t)$ clearly show, Krylov complexity can have qualitatively different behavior than the holographic and computational, 
e.g.~Nielsen, complexity \cite{Susskind:2014rva,Brown:2015bva,Brown:2015lvg,Ben-Ami:2016qex,Chapman:2016hwi,Carmi:2017jqz,Belin:2021bga,caputa2019quantum,caputa2022geometry}. 

Our work raise numerous questions, including  a number of mathematical questions about the relation between $b_n$ and $C(t)$. Among them the features of $C(t)$ which lead to asymptotic persistent staggering \eqref{asymptm} or two slopes \eqref{asymptmS3} behavior and how to deduce $\alpha_1,\alpha_2,c_1,c_2$ from $C(t)$. Here we presume a simple generalization of \eqref{asympt} is possible, though in section \ref{sec:S3S1} we saw the example when the asymptotic of $b_n$ was dependent on the exponentially suppressed contributions to $C(t)$.   
There are many other questions, how to quantitatively relate $\lambda_K$ to $c_1,c_2$ in case of persistent staggering and maximal or time-averaged $K(t)$ to $\alpha_1,\alpha_2$  in case of two slopes behavior. There are also questions about physics of Krylov complexity. Can we prove generalized Maldacena-Shenker-Stanford bound \eqref{MSS} in full generality? Is there any  holographic counterpart of $K(t)$? 
What is the behavior of $b_n$ and $K(t)$ in interacting QFTs? We leave these and other questions for the future.

\section*{\bf Acknowledgments}
We thank Dmitri Trunin for collaboration at the early stages of this project and Luca Delacretaz, Matthew Dodelson, Oleg Lychkovskiy, Julian Sonner,  Subir Sachdev,  Adrian Sanchez-Garrido, and Erez Urbach  for discussions. AA  acknowledges support from a Kavli ENSI fellowship and the NSF under grant number DMR-1918065.
AD is supported by the National Science Foundation under Grants PHY-2013812 and 2310426.

\appendix 
\section{Massive free theories}
\label{app:freetheories}
\subsection{Massive scalar}
We start with a free massive scalar in $\R^d$ and calculate  thermal two-point function 
\bea
 \langle \phi(\tau) \phi(-1/2) \rangle_\beta = \sum_{n=-\infty}^\infty {(-1)^n\over \beta} \int {d^{d-1} k \over (2\pi)^{d-1}} {e^{i 2\pi n \tau} \over \big({2\pi n\over\beta}\big)^2 + k^2 + m^2}. \label{2ptT}
\eea
Here $\tau=it/\beta$ is the Euclidean time. One operator is placed at $\tau=-1/2$, thus this is a Wightman-ordered correlator .  
\bea
\nonumber
 \langle \phi(\tau) \phi(-1/2) \rangle_\beta \propto {\rm Tr}(e^{-\beta H/2}\phi(t,x) e^{-\beta H/2} \phi(0,x)).
\eea
To evaluate \eqref{2ptT} we substitute the sum over $n$ by a contour integral going over the singularities of $ \sinh\big({\tilde k_0\over 2}\big)$,
\bea
 \langle \phi( \tau) \phi(-1/2) \rangle_\beta ={\beta^{2-d}\over 4\pi i } \int_\mathcal{C} d\tilde k_0 \int {d^{d-1} \tilde k \over (2\pi)^{d-1}} 
 {e^{ \tilde k_0  \tau}\over \sinh\big({\tilde k_0\over 2}\big) \big(\tilde m^2 + \tilde k^2 - \tilde k_0^2\big)}.
\eea
The  contour can be deformed and closed through the infinite semicircle at ${\rm Re} \tilde k_0 \rightarrow \infty$, yielding
\bea
\nonumber
C(t)={\rm Tr}(e^{-\beta H/2}\phi(t,x) e^{-\beta H/2} \phi(0,x))\propto {\beta^{2-d} \over (4\pi)^{d-1\over 2}\Gamma({d-1\over 2})} \int_{\tilde{m}}^\infty (y^2-\tilde{m}^2)^{d-3\over 2} {\cosh (y\, \tau )\over \sinh(y/2)} dy.
\eea

\subsection{Massive fermion}
Similarly for the massive fermion 
\bea
 \langle \psi_\alpha^\dagger(\tau) \psi_\sigma (-1/2) \rangle_\beta =  {1 \over \beta} \int {d^{d-1} k \over (2\pi)^{d-1}}  \sum_{n=-\infty}^\infty {(-1)^{n}\over \beta}
 {\big(\pi (2n+1)\delta_{\alpha\sigma} + i \,  \tilde m \,\gamma^0_{\alpha\sigma}\big)e^{i \pi (2n+1) \tau} \over \big({(2n+1)\pi\over\beta}\big)^2 + k^2 + m^2}.
\eea
We work in Dirac representation such that $\gamma^0_{\alpha\sigma} =\pm \delta_{\alpha\sigma}$ and use the same trick with the contour integral 
\bea
 \langle \psi_\alpha^\dagger(\tau) \psi_\sigma (-1/2) \rangle_\beta =  {\beta^{1-d}\over 4\pi i } \int_\mathcal{C} d\tilde k_0 \int {d^{d-1} \tilde k \over (2\pi)^{d-1}} 
 {(\tilde k_0 \pm \tilde m) \, e^{ \tilde k_0  \tau}\over \cosh\big({\tilde k_0\over 2}\big) \big(\tilde k^2 + \tilde m^2  - \tilde k_0^2\big)}.
\eea
After deforming the integration contour we arrive at 
\bea
{\rm Tr}(e^{-\beta H/2}\psi^\dagger (t,x) e^{-\beta H/2} \psi(0,x))\propto 
{\beta^{1-d} \over (4\pi)^{d-1\over 2} \Gamma\big({d-1\over 2}\big)} \int_{|\tilde m|}^\infty dy \, ( y^2 - \tilde m^2)^{d-3\over 2}  
 {y\cosh(y \tau) \pm \tilde m \sinh(y \tau)\over \cosh\big({y\over 2}\big)}. \nonumber
\eea

\section{Lanczos coefficients for general $C(t)$}
\label{app:abn}
In the most general case, when $C(t)$ is not a real-valued even function, besides coefficients $b_n$, Lanczos coefficients also include $a_n$.  Their definition in terms of Krylov basis can be found in e.g.~\cite{dymarsky2020quantum}, and the explicit expression in terms of $C(t)$ is as follows. For the  analytically continued $C(\tau)$, $\tau=it$, we first define  an $n \times n$ 
Hankel matrix of derivatives ${\mathcal M}^{(n)}_{ij}=C^{(i+j)}(\tau)$ and variables $\tau_n={\rm det}\, {\mathcal M}^{(n)}$, $\tau_0\equiv 1$. Then 
\bea
\left. b_n={\tau_{n-1} \tau_{n+1}\over \tau_n^2}\right|_{\tau=0},\\
\left. a_n={d\over d\tau}(\tau_n-\tau_{n-1})\right|_{\tau=0}.
\eea

\section{Free scalar on $\Sp^{d-1} \times \Sp^1$}
\label{app:S3S1}

Let us consider a compact space $\mathbb{S}^{d-1}$ of radius $R$. The thermal correlation function of the scalar field living on $\mathbb{R}\times \mathbb{S}^{d-1}$ can be expressed in terms of the heat kernel $ K_\mathcal{M}(t,x,y)=\langle x|\exp(-t \hat D)|y\rangle$, see {\it e.g., } \cite{Vassilevich:2003xt}
\bea
 \langle \phi(x) \phi(y) \rangle_\beta = \int_0^\infty dt K(t,x,y) e^{-t m^2_\text{eff}}~,
\eea
where $\mathcal{M}=\mathbb{S}^1\times \mathbb{S}^{d-1}$  ($\mathbb{S}^1$ represents a thermal circle parametrized by $\tau$), and $\hat D$ is the second order differential operator of the Laplace-Beltrami type, 
\bea
 \hat D= - {d\over d\tau^2} - \Delta_{\mathbb{S}^{d-1}}~,
\eea
where $\Delta_{\mathbb{S}^{d-1}}$ is the scalar Laplacian on a sphere. Finally $m$ denotes the 'effective' mass of the field under study
\bea
m^2_\text{eff}=  m^2+ \xi \mathcal{R} 
\eea
where  $\xi={d-2\over 4(d-1)}$ is the conformal coupling and $\mathcal{R}=\mathcal{R}(\mathbb{S}^{d-1}) = {(d-1)(d-2)/R^2}$ is the Ricci scalar of the background geometry.

The wave operators are separable on the product manifold $\mathbb{S}^1\times \mathbb{S}^{d-1}$ and hence the heat kernel can be expressed as the product of the two individual heat kernels on $\mathbb{S}^1$ and $\mathbb{S}^{d-1}$, i.e.,
\bea
  K_\mathcal{M} =  K_{\mathbb{S}^1} \, K_{\mathbb{S}^{d-1}}
\eea
$K_{\mathbb{S}^1}$ can be readily evaluated using the method of images. It is given by an infinite sum of the scalar heat kernels on $\mathbb{R}$, which are shifted by integer multiples of $\beta$ with respect to each other to maintain periodic boundary conditions for the scalar field on a circle, namely
\bea
 K_{\mathbb{S}^1}(t,\tau) = {1\over \sqrt{4\pi t}} \sum_{n=-\infty}^\infty e^{-{(\tau + n\beta)^2 \over 4t}}
\eea
In fact, the heat kernel $K_{\mathbb{S}^{d-1}}$ is also  known in full generality \cite{Camporesi:1994ga}, {\it e.g.,}
\bea
K_{\mathbb{S}^{3}}(t,x,x)={e^{t/R^2}\over (4\pi t)^{3/2}} \sum_{n=-\infty}^\infty e^{-{\pi^2 R^2 n^2\over t}} \Big( 1- 2 {\pi^2R^2n^2\over t} \Big)~.
\eea
In $d=4$, we thus get
\bea
 C_\phi(\tau)= \sum_{n,\ell =-\infty}^\infty \int_0^\infty   {dt \, e^{-t \, m^2}\over (4\pi t)^2} e^{-{(\tau + n\beta)^2 + (2 \pi R\ell)^2 \over 4t}} \Big( 1- 2 {\pi^2R^2 \ell^2\over t} \Big)~,
\eea
Integrating over $t$, yields
\begin{eqnarray}
 C_\phi(\tau)&=& m \sum_{n,\ell = -\infty}^\infty   { A \, K_1(Am) - 4 B^2 m \, K_2(Am)\over 4 \pi^2 A^2} ~.
 \nonumber \\
 A^2&=&(\tau + n\beta)^2 + (2 \pi R\ell)^2 \,, ~ B =  \pi R\ell ~.
\end{eqnarray}
For $\beta m, Rm \gg 1$ the sum is dominated by a few terms in the vicinity of $n=\ell=0$. 

\noindent
$C_\phi(\tau)$ simplifies in the case of conformally coupled scalar ($m=0$)
\bea
 C_\phi(\tau)= {1\over 4\pi^2}\sum_{n,\ell =-\infty}^\infty   {(\tau + n\beta)^2 - (2 \pi R\ell)^2 \over\big( (\tau + n\beta)^2 + (2 \pi R\ell)^2\big)^2 } ~.
% -  {2(2 \pi R\ell)^2\over ((\tau + n\beta)^2 + (2 \pi R\ell)^2)^2} \Big)~,
\label{Csphere}
\eea
Upon redefinition $\tau\rightarrow \beta \tau$ and $R\rightarrow \beta R $ we arrive at \eqref{nl}.

\section{Thermal 2pt function in XY model}
\label{app:spinchain}
Consider the Hamiltonian of integrable $XY$ model  with periodic boundary conditions, $j=0$ is the same as $j=N$, 
\begin{gather}
    H = \sum_{j=1}^{N} \left[(1+\gamma) S^x_j S^x_{j+1} + (1-\gamma) S^y_j S^y_{j+1} \right] - h \sum_{j=1}^N S_j^z.
\end{gather}
This chain is diagonalizable by the Jordan-Wigner transform \cite{NIEMEIJER1967377}, with the quasiparticle energies given by $\epsilon_k = \sqrt{(\cos k - h)^2 + \gamma^2 \sin^2 k}$. Here $k$ varies between $0$ to $2 \pi$.
The autocorrelation function at inverse temperature $\beta$ is defined as
\bea
C^{z}_{\beta}(t) \equiv \langle S^z_0 S^z_0(t) \rangle_{\beta}={\rm Tr}(e^{-\beta H} S^z_0 S^z_0(t))/{\rm Tr}(e^{-\beta H}).
\eea
Before giving the expression for $C^z_{\beta}(t)$ we first thermal expectation value 
\bea
m_z(h) = \langle S^z_0\rangle_{\beta} = - \frac{1}{2 \pi} \int_{0}^{\pi}dk \cos\left\{\tan^{-1}\left( \frac{\gamma \sin k}{\cos k - h} \right) \right\} \tanh \frac{\beta \epsilon_k}{2},
\eea
and one can show that $\lim\limits_{h \to 0} m^z(h) = 0$. Finally, the autocorrelation function is given by 
\begin{gather}
C^z_{\beta}(t) = m^2_z(h) + \left[ \frac{1}{2 \pi} \int\limits_0^{\pi} dk \left\{ \cos (\epsilon_k t) + i \sin (\epsilon_k t) \tanh \frac{\beta \epsilon_k}{2} \right\} \right]^2\\
- \left[ \frac{1}{2 \pi} \int\limits_0^{\pi} dk \cos (2\lambda_k) \left\{ i \sin (\epsilon_k t) + \cos (\epsilon_k t) \tanh \frac{\beta \epsilon_k}{2} \right\} \right]^2,\notag
\end{gather}
where $\lambda_k = \frac{1}{2} \tan^{-1} \frac{\gamma \sin k}{\cos k - h}$.

This expression becomes particularly simple, when we set $\gamma = 0$, $h = 0$ which corresponds to isotropic $XY$ model $H = \sum_{j=1}^{N} \left[ S^x_j S^x_{j+1} + S^y_j S^y_{j+1} \right]$,
\bea
C_{\beta}(t) = \frac{1}{4 \pi^2} \left(\int\limits_{-1}^1 \frac{dc}{\sqrt{1 - c^2}} \left[ \cos (tc) - i \sin (tc) \tanh \frac{\beta c}{2} \right] \right)^2,
\eea
which becomes even simpler upon the shift $t \to t - i \beta/2$ that  gives the Whightman-ordered correlator, cf.~with \eqref{XYi},
\bea
C_{\beta}^{W}(t) = \frac{1}{4 \pi^2} \left[\int\limits_{-1}^1 \frac{dc}{\sqrt{1 - c^2}} \frac{\cos (tc)}{\cosh\beta c/2 } \right]^2. 
\eea
 
\section{Dyck paths integral for two slopes scenario}
\label{app:a1a2}
\subsection{Integral over Dyck paths formalism for two branches}
\label{E1}
Here we will generalize the derivation of the asymptotic behavior of moments 
\bea
\label{mu2def}
\mu_{2n}=\int d\omega f^2(\omega) \omega^{2n}
\eea from \cite{Avdoshkin_2020} to the case when $b_n$ form two continuous branches for large $n$, $b_{2n} = b_{even}(2n)$ and $b_{2n+1} = b_{odd}(2n)$. 
We also introduce 
\bea
\label{epsilon}
\epsilon(n)=b_{even}(n)/b_{odd}(n),\\
b(n)=\sqrt{b_{even}(n) b_{odd}(n)},
\label{bdef}
\eea 
and assume they are smooth functions of their argument. 

We start with the Dyck path sum representation of  moments  \cite{Parker_2019}
\bea \label{eq:mu_2n_def}
\mu_{2n} = \sum_{\{h_k\} \in D_n} \prod_{k=0}^{2n-1} b_{(h_k + h_{k+1})/2},
\eea
where $D_n$ is the set of all Dyck paths.

We approximate each Dyck path with $h_i = 1/2 + 2n f(i/2n)$, where $f(t)$ is a continuous function defined on $0 < t < 1$ and satisfying $|f'(t)| < 1$.
Let us consider N consecutive steps in a Dyck path, where $h_{k+1} - h_{k} = 1$ for $n$ distinct values of $k$ so that $f' = \frac{2n - N}{2}$. We additionally require that the index of $b$ is even $d$ times more than it is odd. We split the interval in pairs and we will denote the steps where the value of $h$ increased (decreased) as "+"  or "-" correspondingly.

For pairs "++" or "-~-" one factor of $b$ has an odd index and another has an even one. For pairs "+-" and "-+" both factors will have the same parity that depends on the current value of $h_i$. If we denote the number of "++" pairs as $m$ and the number of opposite sign pairs that lead to two even factors as $l$ we obtain $d = 4l - 2n + 4m$. Assuming our interval is small, its weight in \eqref{eq:mu_2n_def} is given by
\bea
 b_{odd}^{ N/2 - 2l + n - 2m} b_{even}^{ N/2 + 2l - n + 2m} =\sqrt{ b_{odd} b_{even}}^{ N} (b_{even}/b_{odd})^{2l - n + 2m}.
\eea 
The parameters satisfy $0 < n < N$, $n - N/2 <m<n/2$ and $0 < l < n - 2m$. The number of paths for fixed $n$, $m$ and $l$ is given by $C_{N/2}^{m} C_{N/2-m}^{n - 2m} C_{n - 2m}^l$. This product can be approximated with the help of the Stirling's formula
\bea\nonumber 
&& \log C_{N/2}^{m} C_{N/2-m}^{n - 2m} C_{n - 2m}^l \approx\\
&& \frac{N}{2} \left[ S(\alpha) + (1 - \alpha) S\left(\frac{f' +1 -2\alpha}{1- \alpha}\right) + (f' + 1 -2 \alpha) S\left(\frac{\beta}{f' + 1 - 2 \alpha}\right) \right],
\eea
where $n =N\frac{f'+1}{2}$, $\alpha = \frac{m}{N/2}$, $\beta = \frac{l}{N/2}$ and
\bea
S(\alpha) = - \alpha \log \alpha - (1- \alpha) \log (1 - \alpha).
\eea
 We can evaluate the sum over $l$ and $m$
\bea
\sqrt{ b_{odd} b_{even}}^{ N} \sum_{m, l} C_{N/2}^{m} C_{N/2-m}^{n - 2m} C_{n - 2m}^l \epsilon^{2l - n + 2m},
\eea 
where $\epsilon = b_{even}/b_{odd}$ via the saddle point approximation.  This is function \eqref{epsilon}, now understood as a smooth function of $f$,  $\epsilon(2nf)$. The function to be minimized is
\bea
&&W(\alpha, \beta, f', \epsilon) =\nonumber\\
&& S(\alpha) + (1 - \alpha) S\left(\frac{f' +1 -2\alpha}{1- \alpha}\right) + (f' + 1 -2 \alpha) S\left(\frac{\beta}{f' + 1 - 2 \alpha}\right) + (2 \alpha + 2 \beta - f' - 1) \log \epsilon.\nonumber
\eea
The saddle point equations  are
\bea
\epsilon^2 (1 - 2 \alpha - \beta + f')^2 = \alpha (\alpha - f'),\\
\epsilon^2 (1 - 2 \alpha - \beta + f') = \beta.
\eea
It is solved by
\bea
\alpha^* = \frac{f' (1 - \epsilon^2)^2 - 4 \epsilon^2 + (1+\epsilon^2)\sqrt{4 \epsilon^2 + (\epsilon^2 - 1)^2 f'^2}}{2 (\epsilon^2 - 1)^2},\\
\beta^* =\frac{\epsilon^2}{(\epsilon^2 - 1)^2}(1 + \epsilon^2 - \sqrt{4 \epsilon^2 + (\epsilon^2 - 1)^2 f'^2}).
\eea
Plugging this back in the expression for $W$, we obtain
\bea
\nonumber
W_{\text{eff}}(f', \epsilon) = -\frac{f'}{2}\log \frac{\e^{{4}} f'^2+f' (f' + \sqrt{4 \e^2 + (\e^2-1)^2 f'^2})+\e^2 (2 + f' \sqrt{4 \e^2 + (\e^2-1)^2f'^2})}{2 \e^2 (1 - f')^2} +\\ {2\log(1+\e)}
+ \log \frac{(\e - 1)^2}{\e + \e^3 - \e\sqrt{4 \e^2 +(\e^2-1)^2 (f')^2}}.\qquad \qquad 
\label{Weff}
\eea
%\bea
%\nonumber
%W_{\text{eff}}(f', \epsilon) = -\frac{f'}{2}\log \frac{\e^{{4}} f'^2+f' (f' + \sqrt{4 \e^2 + (\e^2-1)^2 f'^2})+\e^2 (2 + f' \sqrt{4 \e^2 + (\e^2-1)f'^2})}{2 \e^2 (1 - f')^2} +\\ {2\log(1+\e)}
%+ \log \frac{(\e^2 - 1)^2}{\e + \e^3 - \e\sqrt{4 \e^2 +(\e^2-1)^2 (f')^2}}.\qquad \qquad 
%\label{Weff}
%\eea
%\bea
%\nonumber
%W_{\text{eff}}(f', \epsilon) = -\frac{f'}{2}\log \frac{\e^{\textcolor{red}{4}} f'^2+f' (f' + \sqrt{4 \e^2 + (\e^2-1)^2 f'^2})+\e^2 (2 + f' \sqrt{4 \e^2 + (\e^2-1)f'^2})}{2 \e^2 (1 - f')^2} +\\ \textcolor{red}{2\log(1+\e)}
%+ \log \frac{(\e^2 - 1)^2}{\e + \e^3 - \e\sqrt{4 \e^2 +(\e^2-1)^2 (f')^2}}.\qquad \qquad 
%\label{Weff}
%\eea

Finally, this gives us a path integral representation of moments \eqref{mu2def} in terms of smooth functions (\ref{epsilon},\ref{bdef})
\bea
\mu_{2n} = \int \mathcal{D} f(t) e^{n \int\limits_{0}^1 dt \left( W_{\text{eff}}(f', \epsilon(2nf)) + 2 \log b(2nf) \right)}.
\eea
We can verify that this path integral reduces to the path integral for a single branch developed in \cite{Avdoshkin_2020} when we set $\epsilon = 1$ ($b_{even} = b_{odd}$). While the second term is singular as $\epsilon \to 0$ it is finite if we carefully take the limit
\bea
\lim\limits_{\epsilon \to 1} W_{\text{eff}}(f', \epsilon) = 2 S\left(\frac{f' + 1}{2}\right).
\eea
Alternatively, the saddle point configuration simplifies to
\bea
\alpha^* = \frac{(1 + f')^2}{4}, ~~ \beta^* = \frac{(1 - f')^2}{4}.
\eea 
and again we obtain $W(\alpha^*, \beta^*, f', 1) = 2 S \left(\frac{f' + 1}{2}\right)$. This agrees with the expected result, since $W$ is multiplied by $n$ (it used to be $2n$ in the single branch case).

\subsection{Evaluation of the path integral for two slopes}
\label{E2}
Here we assume that $\epsilon = \text{const}$ and $\sqrt{b_{odd}(n) b_{even}(n)} = \alpha n$.
The action in this case reads
\bea
I = \int\limits_{0}^1 dt \left( W_{eff}(f', \epsilon) + 2 \log f \right),
\eea
where have neglected the constant term $2 \log ( 2 \alpha  n)$ in the integral since it does not affect the shape of the saddle point.
Varying $I$ with respect to $f(t)$, we obtain the EOMs
\bea
-\frac{2}{f(t)} = \frac{f''}{1 - (f')^2} \frac{(1 + \epsilon^2 + \sqrt{
    4 \epsilon^2 + (-1 + \epsilon^2)^2 (f')^2 })}{ \sqrt{
  4 \epsilon^2 + (-1 + \epsilon^2)^2 (f')^2}}.
\eea
%
%\iffalse
%If we take the limit $\epsilon \to 1$, once again, we reproduce the old equation of the saddle point
%\bea
%-\frac{2}{f(t)} = \frac{f''}{1 - (f')^2}.
%\eea
%This equation is solved by
%\bea
%f(t) = \frac{\sin (\pi t)}{\pi}.
%\eea
%The solution of the equation for arbitrary $\epsilon$ is given implicitly by
%
%\bea
%\frac{1}{c_1} F(c_1 f) = t + c_2,\\
%F(x) = \frac{(1 - \epsilon)\sqrt{(1+\epsilon)^2 - x^2}\sqrt{(1-\epsilon)^2 - x^2} F\left(\arcsin\left(\frac{x}{1 - \epsilon}\right), \frac{(1 - \epsilon)^2}{(1 + \epsilon)^2} \right)}{\sqrt{(1-\epsilon^2)^2- 2 (1 + \epsilon^2)x^2 + x^4}}.
%\eea
%The constant $c_2$ can be determined from
%
%\bea
%\frac{1}{c_1}F(0) = 0 = c_2.
%\eea
%Additionally, at the maximum point $t=1/2$ we have
%\bea
% F'(c_1 f(1/2)) = 1/f'.
%\eea
%Thus $F'(x)$ diverges at $c_1 f(1/2)$
%
%\bea
%F'(x) = \frac{1 - \epsilon^2}{\sqrt{\epsilon^4 + (1 - x^2)^2 - 2 \epsilon^2 (1 + x^2)}}.
%\eea
%This function diverges at $x = 1- \epsilon$. Thus, we have $1 - \epsilon = c_1 f(1/2)$.
%
%\fi
We can lower the order of equation by defining $g(f) = f'$:
\bea
\frac{(1 + \epsilon^2 + \sqrt{4 \epsilon^2 + (1 - \epsilon^2)^2 g^2}) g' g}{(1 - g^2) \sqrt{4 \epsilon^2 + (1 - \epsilon^2)^2 g^2}} = -\frac{2}{f}.
\eea
And the solution is
\bea \label{eq:1st_order_eq}
g^2 = \frac{
 \epsilon^4 + (1 - (c_1 f)^2)^2 - 
  2 \epsilon^2 (1 + (c_1 f)^2)}{(1 - \epsilon^2)^2},
\eea
where $c_1$ is some constant. By requiring $g(f(1/2)=f_0) =0$, we arrive at 
\bea
c_1 f_0 = 1 - \epsilon, ~ c_1 = \frac{1- \epsilon}{f_0}.
\eea
Having found $g$ as a function of $f$, we can now solve for $f$ as a function of $t$. Setting $g = f'$ leads to
\bea
\frac{1}{c_1} H(c_1 f, \epsilon) = \frac{t}{1 - \epsilon^2} + c_2,
\eea
where
\bea
H(x, \epsilon) = \frac{(1 + \epsilon) \sqrt{1 - \left(\frac{x}{1 - \epsilon}\right)^2} \sqrt{
   1 - \left(\frac{ x}{1 + \epsilon}\right)^2}
    F\left(
    \arcsin(\frac{x}{1 + \epsilon}), \left(\frac{1 + \epsilon}{1 - \epsilon}\right)^2\right)}{\sqrt{
 \epsilon^4 + (1 - x^2)^2 - 2 \epsilon^2 (1 + x^2)}},
\eea
and $F(x, a)$ is the incomplete elliptic integral (we use the same conventions  as Wolfram Mathematica).
Taking the limit $x \to 1 - \epsilon$ we have
\bea
H(1- \epsilon, \epsilon) = \frac{K\left[\left(\frac{1 - \epsilon}{1 + \epsilon}\right)^2\right]}{1 + \epsilon},
\eea
where $K$ is the complete elliptic integral.

We can use this expression to find $c_1$ and the value of the function at the maximum ($t = 1/2$)
\bea
2 (1 - \epsilon) K\left[\left(\frac{1 - \epsilon}{1 + \epsilon}\right)^2\right] = c_1, ~
f_0 = \frac{1}{2 K\left[\left(\frac{1 - \epsilon}{1 + \epsilon}\right)^2\right]}.
\eea
As a consistency check we verify that for $\epsilon = 1$ we reproduce the old result $\lim\limits_{\epsilon \to 1}f_0 = 1 / \pi$ (the solution for a sinlge branch was $\frac{\sin(\pi t)}{\pi}$).\\

\begin{figure}
    \centering
    \includegraphics[width=0.5\linewidth]{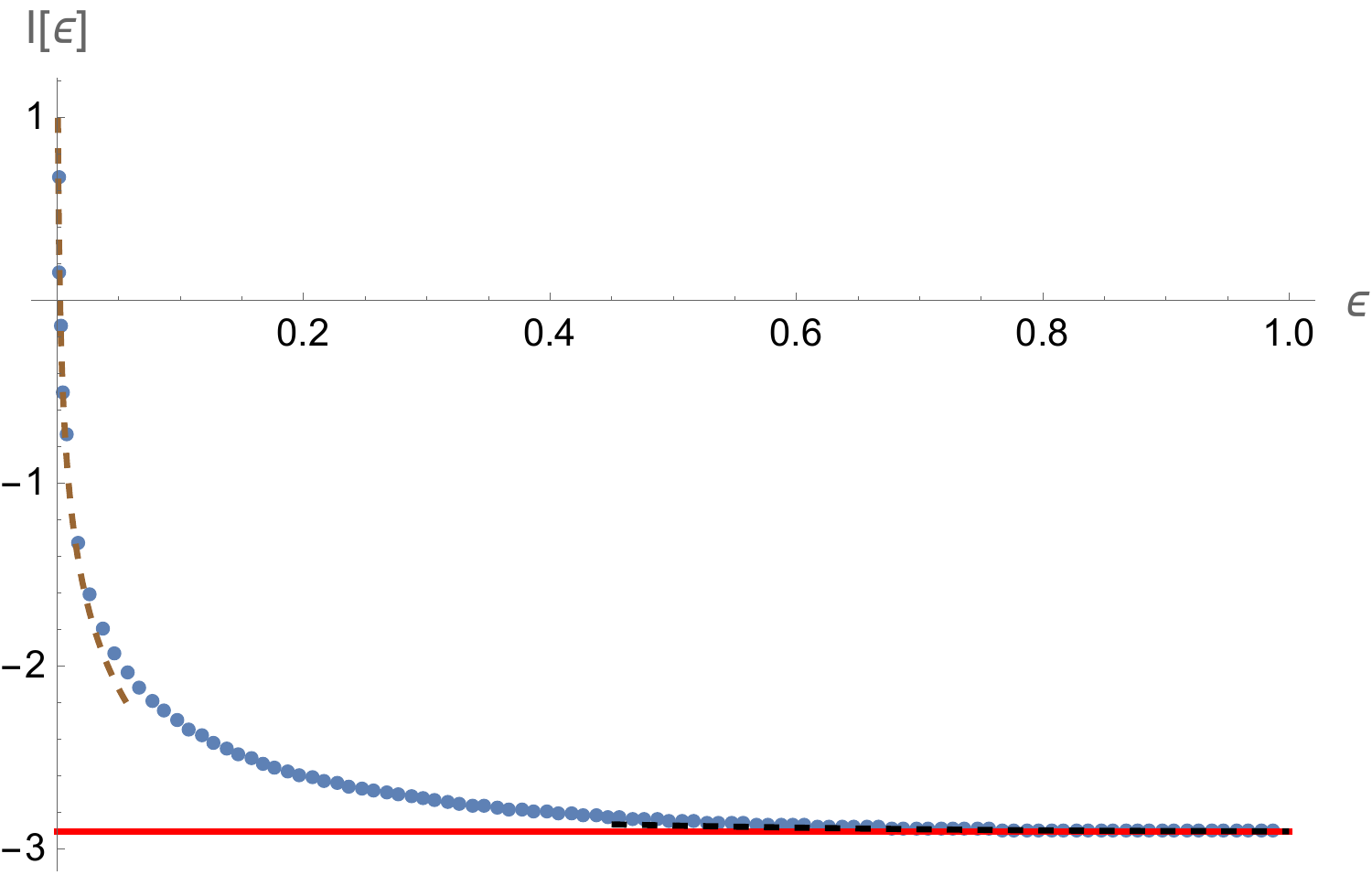}
    \caption{Numeric values of $I(\epsilon)$, shown in blue, vs the asymptotic value $I(1)=2\log \frac{2}{\pi e}$ for $\epsilon=1$ (analytic value for the single branch), in red. Also, the asymptotic  behavior $I(\epsilon)=2\log \frac{2}{\pi e}+(1-\epsilon)^2/8$  for large  $\epsilon$
\eqref{largeeps} (shown in dashed black), and the asymptotic behavior \eqref{smalleps} for small $\epsilon$ (shown in dashed brown).}
    \label{fig:I_plot}
\end{figure}

The ``on-shell'' action can be rewritten as
\bea
I(\epsilon) = 2 \int\limits_{0}^{f_0} \frac{df}{g(f)}\left(W(g(f),\epsilon) + 2 \log f \right),
\eea
where $f_0$ and $g(f)$ are given by
\bea
f_0 &=& \frac{1}{2 K\left[\left(\frac{1 - \epsilon}{1 + \epsilon}\right)^2\right]},\\
g^2(f) &=& \frac{(f^2 - f^2_0) }{(1 + 
   \epsilon)^2 f_0^4}\left[(1 - \epsilon)^2 f^2 - (1 + \epsilon)^2 f_0^2\right].
\eea
Written explicitly, this integral is rather cumbersome and  difficult to deal with. The numeric plot of $I$ as a function of $\epsilon$ is shown in Fig.~\ref{fig:I_plot}. 
The asymptotic behavior of the moments is
\bea
\boxed{\log \mu_{2n} \approx n I(\epsilon) + 2n \log (2 \alpha n).}
\eea
If $\epsilon = 1$, $I(1) = 2\log \frac{2}{\pi e}$ and $\log \mu_{2n} \approx 2n \log \frac{4 \alpha n}{\pi e} $.

\begin{figure}
    \centering
    \includegraphics[width=0.5\linewidth]{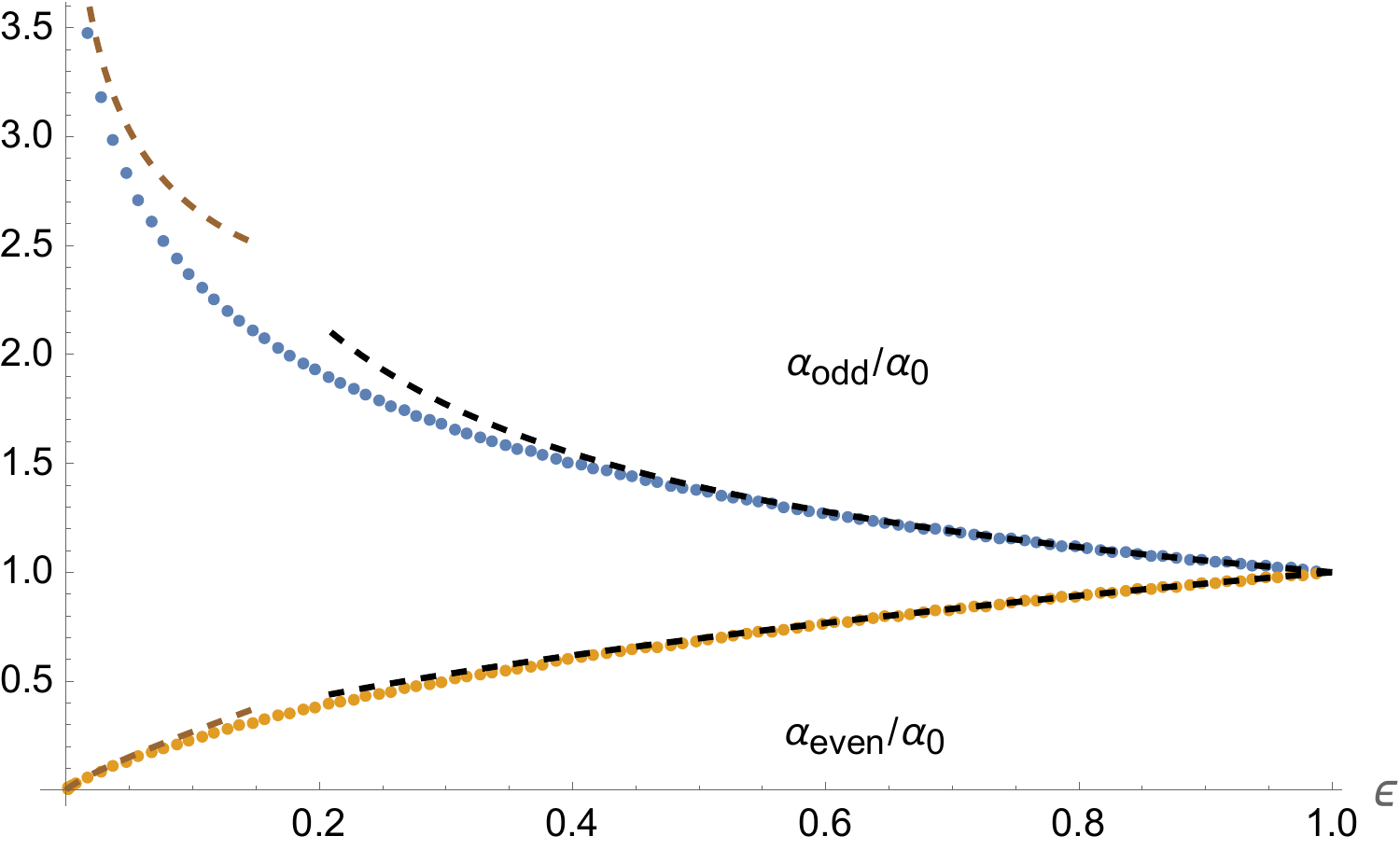}
    \caption{Ratio of $\alpha_{\rm odd}/\alpha_0$ (blue) and $\alpha_{\rm even}/\alpha_0$ (orange) as a function of $\epsilon$, superimposed with the  asymptotic behavior (\ref{oddevensmall1},\ref{oddevensmall2}) for 
    $\epsilon\rightarrow 1$ (dashed black), and (\ref{osmall},\ref{esmall}) for $\epsilon \rightarrow 0$ (dashed brown).}
    \label{fig:oddeven}
\end{figure}

Leading asymptotic behavior of $\mu_{2n}$ for large $n$ is fully determined by the location of the singularity of the correlation function $C(t)$ along the imaginary axis $t=i\tau^*$,
\bea
\log \mu_{2n} \approx 2n \log \frac{4 \alpha_0 n}{\pi e},\qquad  \alpha_0=\pi/(2\tau^*).
\eea 
As we discussed in section \ref{sec:prel}, in field theory, symmetrically ordered correlation function \eqref{C} always has the singularity at $\tau^*=\beta/2$, and $\alpha_0=\pi/\beta$. In the scenario of two slopes, we can express $\alpha_{\rm odd}, \alpha_{\rm even}$ \eqref{asymptmS3} in terms of $\alpha_0$ and $\epsilon$
\bea
{\alpha_{\rm odd} \over \alpha_0}&=& {2 e^{-I(\epsilon)/2}\over \pi e }{1\over \sqrt{\epsilon}},\\
{\alpha_{\rm even} \over \alpha_0}&=& {2 e^{-I(\epsilon)/2}\over \pi e } \sqrt{\epsilon}.
\eea
In particular, plotting $\alpha_{\rm odd}/ \alpha_0$ and $\alpha_{\rm even}/\alpha_0$ numerically, as a function of $\epsilon$, see Fig.~\ref{fig:oddeven}, we find that 
\bea
\label{alphaineq}
\alpha_{\rm odd}\geq \alpha_0\geq \alpha_{\rm even}
\eea
always holds. 

\subsection{Small $\epsilon$ expansion}
In this subsection, we continue discussing the case of two linear branches with different slopes and  will determine the small $\epsilon$ behavior of $I(\epsilon)$. First of all, since we know the exact value of $f(1/2)$ we can find the limit
\bea
f(1/2) = f_0 \approx \frac{1}{|\log \epsilon|}.
\eea
Next, at small $\epsilon$ \eqref{eq:1st_order_eq} simplifies to
\bea
f'(t) = 1 - (f(t)/f_0)^2,
\eea
which is solved by $f(t) = f_0 \tanh(t/f_0)$. While this solution seems to not be symmetric under $t \to 1 - t$, it should be thought of as only applicable at $t < 1/2$ and defined by symmetry for $t>1/2$. This function is essentially constant for $t \gg \frac{1}{|\log \epsilon|}$ and, thus, stitching it with a different function at $t=1/2$ is appropriate.

Now, we expand $W(f', \epsilon)$.
\bea
W(f', \epsilon) \approx S(f') + |\log \epsilon| (1 - f'),\\
I(\epsilon) = 2\int\limits_0^{1/2} dt (S(f') - |\log \epsilon| f') + |\log \epsilon | + 4\int\limits_0^{1/2} dt \log(f) ,\\
\int\limits_0^{\infty} (S(f') - |\log \epsilon| f') \approx \frac{ \pi^2/4 - \log(4)}{ |\log \epsilon|} - 1,\\
\int\limits_0^{1/2} dt \log(f) \approx - \frac{\pi^2}{8} \frac{1}{|\log \epsilon|}   - \frac{1}{2} \log |\log \epsilon|.
\eea
Collecting all the terms together,
\bea
\label{smalleps}
I(\epsilon) = |\log \epsilon| - 2 \log |\log \epsilon| - 2 - \frac{2 \log 4}{|\log \epsilon|} + \cdots.
\eea
Since  the leading behavior of moments is
\bea
\log \mu_{2n} \approx n I(\epsilon) + 2n \log (2 \alpha n),
\eea
this means corresponding correlation function $C(t)$ will have a singularity at 
\bea
it=\tau^*\approx -{\epsilon^{1/2}\log\epsilon\over \alpha} \label{pole}
\eea 
and 
\bea
\label{osmall}
{\alpha_{\rm odd}\over \alpha_0}&=&-{2\over \pi}\log \epsilon\,  e^{-\log(4)/\log\epsilon},\\
{\alpha_{\rm even}\over \alpha_0}&=&-{2\over \pi} \epsilon \log \epsilon\, e^{-\log(4)/\log\epsilon}.
\label{esmall}
\eea
We can check this result by comparing with the small radius expansion \eqref{smallR}, when, at leading order 
\bea
\alpha={2\pi \over \beta R} e^{-{\pi \over 4R}}\,\qquad \epsilon={4e^{-{\pi \over 2R}}},\quad R\rightarrow 0, 
\eea
and we restored $\beta$-dependence. Plugging this into \eqref{pole} yields $\tau^*=\beta/2$, which is the universal result for  field theory and readily follows from \eqref{nl}. In other words, small radius expansion \eqref{smallR} is in agreement with 
\eqref{pole}.

\subsection{$1 - \epsilon$ expansion}
\label{E4}
Let us define $\epsilon = 1 - \delta$. We can expand
\bea
f_0 \approx \frac{1}{\pi} - \frac{\delta^2}{16 \pi}.
\eea
The correction to the action is
\bea\label{eq:delta_S}
\delta S = \frac{\delta^2}{4}\int\limits_{0}^{1} dt (1 - (f')^2).
\eea
Since the variation of the action vanishes on the saddle point solution the contribution coming from the change in the saddle point will be of order $\delta^4$. Thus, to leading order we only need to evaluate the value of \eqref{eq:delta_S},
\bea
\label{largeeps}
\delta S = \frac{\delta^2}{8}.
\eea
Finally,
\bea
\log \mu_{2n} = 2n \left( \log \frac{4 \alpha n}{\pi e} + \frac{\delta^2}{16}\right) + O(\delta^4).
\eea
This means, for $\delta \ll 1$, the correlation function will have  a singularity at $it=\tau^*=\pi/(2\alpha) e^{-\delta^2/16}$. For field theory, in full generality we expect this to be equal to $\tau^*=\beta/2$, and hence $\alpha=\pi/\beta e^{-\delta^2/16}$. This of course reduces to standard result $b_n\sim {\pi\over \beta} n$
when there is one linear slope, as in conformal field theories in flat space considered in \cite{Dymarsky:2021bjq}. 
For the  ``two slopes behavior'' \eqref{asymptmS3} we  find in full generality , 
\bea
\label{oddevensmall1}
\alpha_{\rm odd}=\alpha_0  (1-\delta)^{-1/2} e^{-\delta^2/16},\\
\alpha_{\rm even}=\alpha_0 (1-\delta)^{+1/2} e^{-\delta^2/16},
\label{oddevensmall2}
\eea
when $\delta \ll 1$, i.e.~when $\alpha_{\rm even}/\alpha_{\rm odd}$ is very close to one. 
It would be interesting to compare this prediction with the large $R$ limit of $b_n$ for \eqref{nl}.

\subsection{Persistent staggering behavior}
\label{PSE}
In section \ref{sec:freemassive} we saw that for massive theories Lanczos coefficients exhibit ``persistent staggering'' behavior \eqref{asymptm}. It also can be described via the general formalism developed in section \ref{E1} above. We assume $b_{even}=\alpha n+c_e, b_{odd}=\alpha n+c_o$ and
in this case $\epsilon=1+c/n+\dots$ where $c=c_{o}-c_{e}$ is a constant. 
For $c = 0$ the problem reduces to the original single-slope, no intercept problem, which is solved by $f(t) = \frac{\sin (\pi t)}{\pi}$. For non-zero $c$ we treat $\frac{c}{n}$ as a perturbation and obtain
\bea
W(f', \epsilon) = (\text{c=0 part}) + c^2\frac{\sin(\pi t)^2}{4 n^2} + \dots
\eea
Upon integration, the leading-order correction is $\frac{c^2}{8 n}$, while the $c=0$ result with  $c_{e}=c_{o}\neq 0$ was calculated in \cite{Dymarsky:2021bjq},
\bea
\label{sinbeh}
\log \mu_{2n} = 2n \log \frac{4 \alpha n}{\pi e}+{c_{e}+c_{o}\over \alpha}\log (2n) + \cdots
\eea
It is clear that  $\frac{c^2}{8 n}$ is subleading and does not affect the singularity behavior of $C(t)\sim (t-i\tau^*)^{-2\Delta}$, which follows from \eqref{sinbeh},
\bea
\tau^*={\pi \over 2\alpha},\qquad 2\Delta-1={c_{e}+c_{o}\over \alpha}.
\eea

\bibliographystyle{JHEP}
\bibliography{Krylov}
\end{document}